\documentclass[11pt,letterpaper]{article}

\usepackage{jheppub}
\usepackage{xcolor}
\usepackage{graphicx}
\usepackage{xspace}
\usepackage{wrapfig,enumerate,slashed}
\usepackage{subfigure}
\usepackage{footmisc}
\usepackage{amsmath}
\usepackage{wasysym} 
\usepackage{graphicx}
\usepackage{color}
\usepackage{orcidlink}
\usepackage{comment}
\usepackage{hyperref}
\usepackage{booktabs}
\usepackage{multirow}
\usepackage{cleveref}

\renewcommand{\vec}{\bf}

\newcommand{\SM}{\text{SM}}
\newcommand{\NN}{\text{NN}}

\newcommand{\OHBtil}{\ensuremath{\widetilde{\mathcal{O}}_{\Phi\widetilde{B}}}}
\newcommand{\OHWtil}{\ensuremath{\widetilde{\mathcal{O}}_{\Phi\widetilde{W}}}}
\newcommand{\OHWtilB}{\ensuremath{\widetilde{\mathcal{O}}_{\Phi\widetilde{W}B}}}
\newcommand{\cHBtil}{\ensuremath{c_{\Phi\widetilde{B}}}}
\newcommand{\cHWtil}{\ensuremath{c_{\Phi\widetilde{W}}}}
\newcommand{\cHWtilB}{\ensuremath{c_{\Phi\widetilde{W}B}}}

\newcommand{\mg}{{\sc MadGraph5}}
\newcommand{\delphes}{{\sc Delphes}}
\newcommand{\Htautau}{\ensuremath{H \rightarrow \tau^{+}\tau^{-}}\xspace}

\begin{document}
\allowdisplaybreaks
\flushbottom
%%%%%%%%%%%%%%%%%%%%%%%%%%%%%%%%%%%
\title{Future Collider Perspectives on Higgs CP Violation}
%%%%%%%%%%%%%%%%%%%%%%%%%%%%%%%%%%%
\author[a]{Arun Atwal,}
\author[b]{Jessica Burridge\orcidlink{0009-0005-3189-2958},}
\author[b]{António Jacques Costa\orcidlink{0000-0001-6305-8400},}
\author[b]{Christoph Englert\orcidlink{0000-0003-2201-0667},}
\author[c,d]{Sinead Farrington\orcidlink{0000-0001-5350-9271},}
\author[e]{Jay Nesbitt\orcidlink{0009-0000-7954-1640},}
\author[f]{Leonor Santos Pereira Trigo\orcidlink{0009-0009-4896-9455},}
\author[b]{Andrew Pilkington\orcidlink{0000-0001-8007-0778},}
\author[e]{Aidan Robson\orcidlink{0000-0002-1659-8284},}
\author[c]{Júlia Cardoso Silva\orcidlink{0000-0002-5987-2984},}
\author[a]{Sarah Williams\orcidlink{0000-0001-6174-401X},}
\author[e]{Yuyang Zhang\orcidlink{0009-0002-3419-5814}}

\affiliation[a]{Cavendish Laboratory, University of Cambridge, J.~J.~Thomson Avenue,\\Cambridge, CB3 0US, United Kingdom}
\affiliation[b]{Department of Physics \& Astronomy, University of Manchester, Oxford Road,\\Manchester M13 9PL, United Kingdom}
\affiliation[c]{School of Physics \& Astronomy, University of Edinburgh, Peter Guthrie Tait Road,\\Edinburgh EH9 3FD, United Kingdom}
\affiliation[d]{Particle Physics Department, Rutherford Appleton Laboratory, Harwell Campus,\\Didcot, OX11 0QX, United Kingdom}
\affiliation[e]{School of Physics \& Astronomy, University of Glasgow, University Avenue,\\Glasgow G12 8QQ, United Kingdom}
\affiliation[f]{Deutsches Elektronen-Synchrotron DESY, Platanenallee 6, 15738 Zeuthen, Germany}

%%%%%%%%%%%%%%%%%%%%%%%%%%%%%%%%%%%
\emailAdd{jessica.burridge@postgrad.manchester.ac.uk}
\emailAdd{antonio.mendesjacquesdacosta@manchester.ac.uk}
\emailAdd{christoph.englert@manchester.ac.uk}
\emailAdd{sinead.farrington@ed.ac.uk}
\emailAdd{jay.nesbitt@glasgow.ac.uk}
\emailAdd{leonor.santos@desy.de}
\emailAdd{andrew.pilkington@manchester.ac.uk}
\emailAdd{aidan.robson@glasgow.ac.uk}
\emailAdd{jcardoso@ed.ac.uk}
\emailAdd{sarahw@hep.phy.cam.ac.uk}
\emailAdd{y.zhang.15@research.gla.ac.uk}
%%%%%%%%%%%%%%%%%%%%%%%%%%%%%%%%%%%

\abstract{The search for new sources of CP violation is a cornerstone of the beyond the Standard Model phenomenology programme at the LHC and beyond. We provide a comprehensive analysis of such searches at a range of future facilities with the aim of informing the currently unfolding future collider roadmap. Focussing on new sources of CP violation specifically in the gauge-Higgs sector, we demonstrate the outstanding potential held by future electron-positron and proton-proton colliders to reveal and identify BSM physics with direct relevance for the observed matter-antimatter asymmetry. In particular, the future colliders will provide an order of magnitude improvement in sensitivity to anomalous CP-violating interactions induced by dimension-six effective field theory operators when compared to the high-luminosity LHC programme.}

\preprint{}

\maketitle

%%%%%%%%%%%%%%%%%%%%%%%%%%%%%%%%%%%%%%%%%%%%%%%%%%%%%
\section{Introduction}
\label{sec:intro}
%%%%%%%%%%%%%%%%%%%%%%%%%%%%%%%%%%%%%%%%%%%%%%%%%%%%%
There is compelling experimental evidence indicating that the Standard Model (SM) of Particle Physics is incomplete. Beyond several theoretical fine-tuning issues, the SM fails to account for the observed matter-antimatter asymmetry in the visible Universe, providing strong motivation for physics beyond the Standard Model (BSM). This shortcoming is formally framed by the Sakharov criteria~\cite{Sakharov:1967dj}, which, from an SM perspective, necessitate additional sources of CP violation and BSM out-of-equilibrium dynamics.

The absence of direct evidence for BSM physics in current searches deepens the puzzle of the SM’s incompleteness. A contemporary way to navigate this impasse is through Effective Field Theory~\cite{Weinberg:1978kz} (EFT), which interprets the lack of direct new physics signals as indicative of a mass gap between the energy scales currently accessible at the Large Hadron Collider (LHC) and the intrinsic scale of new physics. Within this framework, new physics can be probed {\emph{agnostically}} by identifying deviations from SM interactions when the high-energy behaviour of SM fields and symmetries is experimentally scrutinised.

At the leading order in the expansion of the generic new physics away from the Standard Model relevant for collider physics, such modifications are given by dimension-six operators~${\cal{O}}_i$~\cite{Grzadkowski:2010es}
\begin{equation}
{\cal{L}}_{\text{BSM}} = {\cal{L}}_{\text{SM}} + \sum_i {c_i \over \Lambda^2} {\cal{O}}_i ,
\end{equation} 
where the Wilson coefficient $c_i$ parametrises the strength of the new physics interaction at the new physics scale $\Lambda$. In this parametrisation, CP violation in the weak bosonic gauge-Higgs sector takes a comparably compact form; relevant interactions are given by\footnote{Assuming a vanishing weak CP phase of heavy fermions, these operators form a closed set under the dimension-six renormalisation group flow~\cite{Grojean:2013kd,Jenkins:2013zja,Jenkins:2013wua,Alonso:2013hga,Englert:2014cva}. In principle, we could further consider the gluonic $\Phi^{\dagger}\Phi G^{\mu\nu}\Phi \widetilde{G}_{\mu\nu}$, which, however, does not enable a reasonable comparison between lepton and hadron machines. Phenomenological analyses on the latter for the HL-LHC, and ways to disentangle it from competing top-Yukawa modifications have been presented elsewhere~\cite{Englert:2019xhk}.}
\begin{equation}
\label{eq:ops}
\begin{split}
\OHBtil &= \Phi^{\dagger}\Phi B^{\mu\nu}\Phi \widetilde{B}_{\mu\nu},\\
\OHWtil &= \Phi^{\dagger}\Phi W^{i\,\mu\nu}\widetilde{W}^i_{\mu\nu},\\
\OHWtilB &= \Phi^{\dagger}\sigma^i \widetilde{W}^{i\,\mu\nu}B_{\mu\nu},
\end{split}
\end{equation}
where $\Phi$ is the Higgs doublet and the $\widetilde{X}_{\mu\nu}=\epsilon_{\mu\nu\rho\delta} X^{\rho \delta}/2$ is the dual field strength tensor for the $X=B,W$ hypercharge and weak gauge fields, respectively. The $\sigma^i$ are the Pauli matrices.

Considering only these dimension-six operators\footnote{It is further worth highlighting that a range of lepton-sector extensions have been proposed in the literature~\cite{Bakshi:2021ofj,DasBakshi:2020ejz} that source these interactions {\emph{dominantly}} to address recently observed anomalies in electroweak ATLAS data~\cite{ATLAS:2020nzk}.}, any scattering amplitude can be expanded as a sum of SM and dimension-six parts, ${\cal{M}}={\cal{M}}_{\text{SM}} + {\cal{M}}_\text{d6}/\Lambda^2$, leading to
\begin{equation}
|\mathcal{M}|^2 = |\mathcal{M}_{\text{SM}}|^2 + {2\over \Lambda^2}\, \text{Re}(\mathcal{M}^*_{\text{SM}}\mathcal{M}_{\text{d6}}) + {1\over \Lambda^4}\,|\mathcal{M}_{\text{d6}}|^2  
\end{equation}
for differential probabilities. The interference term $\text{Re}(\mathcal{M}^*_{\text{SM}}\mathcal{M}_{\text{d6}})$ is the leading correction to the SM when experimental sensitivity is high enough to establish robust constraints. For the CP-odd interactions noted above, the interference term is CP-odd and thus produces asymmetries in CP-odd collider observables whilst integrating to zero for any CP-even observable. The `squared' dimension-6 contributions, however, produce CP-even modifications to the scattering rate and produce no asymmetries. CP-odd observables targeting the interference term are therefore needed if the goal is to set {\emph{genuine}} limits on CP-violating effects in the gauge-Higgs interactions. 

The LHC is expected to increase its sensitivity to these interactions during its high-luminosity (HL) phase as they leave tell-tale signatures in CP-sensitive observables in relatively clean, albeit statistics-limited, final states such as $H\to ZZ^\ast, WW^\ast$, weak boson fusion (WBF) Higgs production, and weak boson pair production. In particular, the use of highly sensitive observables will qualitatively modify earlier sensitivity estimates, which have served as a baseline for informing future collider cases. In light of this progress, it is prudent to reconsider future collider cases that are now under active discussion in both the ECFA process and the imminent update to the European Strategy for Particle Physics (ESPPU). This is the purpose of this work. 

We structure this paper as follows. In Section~\ref{sec:setup}, we introduce the collider cases that we consider in this study, namely the HL-LHC, FCC-ee, FCC-hh, and a linear collider facility (LCF). We also provide a brief overview of our simulation setup for each collider. In Section~\ref{sec:cp_obvs} we introduce the CP-sensitive observables that are capable of projecting out the interference contributions to the squared scattering amplitude. The limit setting procedure is described in Sec.~\ref{sec:limits}. Sections~\ref{sec:eezh}-\ref{sec:vbfh} contain the results of our studies, covering a variety of different Higgs boson production/decay channels at different colliders. We summarise and conclude in~Sec.~\ref{sec:conc}. 

%%%%%%%%%%%%%%%%%%%%%%%%%%%%%%%%%%%%%%%%%%%%%%%%%%%%%
\section{Monte Carlo simulation and collider scenarios}
\label{sec:setup}
%%%%%%%%%%%%%%%%%%%%%%%%%%%%%%%%%%%%%%%%%%%%%%%%%%%%%
This study considers both proton-proton collisions in the context of the HL-LHC and FCC-hh, as well as electron-positron collisions in the context of FCC-ee and LCF. The HL-LHC is expected to begin operation in the 2030s, following extensive upgrades to the accelerator and experiments. HL-LHC will deliver 3~ab$^{-1}$ of $pp$ collisions to the ATLAS and CMS experiments by the end of operation in 2041. The Future Circular Collider (FCC) project~\cite{FCC:2018byv} is a proposal to explore the high-energy frontier after HL-LHC and will be actively discussed as part of the 2026 ESPPU. FCC would involve a new 91~km tunnel being built at CERN. The first stage of FCC would involve $e^{+}e^{-}$ collisions (FCC-ee) at a range of centre-of-mass energies, including at the $Z$-pole ($\sqrt{s}=91$~GeV), the $WW$ threshold ($\sqrt{s}=160$~GeV), the $ZH$ threshold ($\sqrt{s}=240$~GeV), and a high-energy run sensitive to $t\bar{t}$ production ($\sqrt{s}=365$~GeV). The luminosity delivered by FCC-ee at the $ZH$ threshold is expected to be 10.8~ab$^{-1}$. The second stage of FCC would involve proton-proton collisions (FCC-hh) at a centre-of-mass energy of $\sqrt{s}=84$~TeV with an integrated luminosity of 30~ab$^{-1}$.\footnote{The baseline energy of FCC-hh has been updated recently as part of the FCC feasibility study~\cite{Benedikt:2023ayy}. However, the FCC-hh studies carried out in this article predated this update and our analysis is carried out using samples at $\sqrt{s}=100$~TeV (corresponding to the parameters as were used for the FCC Conceptual Design Report~\cite{FCC:2018byv}). The impact of this change in centre-of-mass energy will be a slight reduction in sensitivity, as the Higgs boson production cross sections (and hence event rates) are slightly smaller at $\sqrt{s}=100$~TeV.} The LCF is expected to deliver $e^{+}e^{-}$ collisions at a range of centre-of-mass energies, including at the $Z$-pole ($\sqrt{s}=91$~GeV), the $ZH$ threshold ($\sqrt{s}=250$~GeV), as well as high-energy runs at $\sqrt{s}=350$~GeV, $\sqrt{s}=550$~GeV and $\sqrt{s}=1$~TeV. The LCF will also use polarised beams with nominal beam polarisations at most centre-of-mass energies of 80\% for the electron beam and 30\% for the positron beam. Each beam can also be configured to be majority left-handed or majority right-handed polarised, thus producing four possible initial-state beam polarisations. At the $ZH$ threshold, the fraction of data collected for each $\{e^+, e^-\}$ beam polarisation configuration is expected to be 45\% for both the $\{+80\%,-30\%\}$ and $\{-80\%,+30\%\}$ configurations and 5\% for both of the $\{+80\%,+30\%\}$ and $\{-80\%,-30\%\}$ configurations. The total luminosity at LCF at the $ZH$ threshold is expected to be 3~ab$^{-1}$.

At FCC-ee and LCF, $ZH$ production is studied at the $ZH$ threshold using inclusive decays of the Higgs boson and also in the $H\rightarrow b\bar{b}$ decay channel. $ZH$ production is also considered at the $pp$ colliders (HL-LHC and FCC-hh), along with $H\rightarrow 4l$ production and  WBF $Hjj$ production in the $H\rightarrow \tau\tau$ decay channel. Events are generated using \textsc{MadGraph5\_aMC@NLO}~\cite{Alwall:2014hca}, at leading order accuracy in perturbative QCD. Parton shower, hadronisation and underlying event activity effects are modelled through interfacing to \textsc{Pythia8} \cite{Sjostrand:2007gs, Sjostrand:2014zea}. \textsc{SMEFTSim 3.0}~\cite{Brivio:2017btx, Brivio:2020onw} is used to model the anomalous interactions from the EFT operators introduced in Eq.~\eqref{eq:ops}. For the processes considered, we simulate both the SM and interference terms ($\text{Re}(\mathcal{M}^*_{\text{SM}}\mathcal{M}_{\text{d6}})$), with the interference contributions produced at $c/\Lambda^{2}=1/\textup{TeV}^{2}$. Normalisation factors are applied to cover missing higher-order effects. For $pp$ collisions, cross-checks on the predicted SM signal and background yields are made with relevant LHC analyses at $\sqrt{s}=13$~TeV.

Detector effects are modelled using \textsc{Delphes v3}~\cite{deFavereau:2013fsa}, which provides a parametrised fast simulation of a generic multipurpose detector. For $pp$ collisions at the (HL-)LHC, the default ATLAS card of \textsc{Delphes v3} is used. The {\tt IDEA} detector card provided by \textsc{Delphes v3} is used for FCC-ee, and the default FCC-hh detector card is used for $pp$ collisions.  These correspond to the same detector settings as were used for the FCC CDR~\cite{FCC:2018byv}. At LCF, the {\tt{ILCgen}} card provided by \textsc{Delphes v3} is used. 

%%%%%%%%%%%%%%%%%%%%%%%%%%%%%%%%%%%%%%%%%%%%%%%%%%%%%
\section{CP-sensitive observables}
\label{sec:cp_obvs}
%%%%%%%%%%%%%%%%%%%%%%%%%%%%%%%%%%%%%%%%%%%%%%%%%%%%%
As mentioned above, the CP-odd operator structures considered in this work can be accessed through CP-odd collider observables, directly projecting out the experimental sensitivity on the associated Wilson coefficient in ways that minimise contributions from CP-even interactions. The study of (optimised) asymmetries, therefore, gives rise to an experimentally robust and theoretically appealing approach to limit or discover CP violation: dominant backgrounds will exhibit symmetric distributions and, hence, have a highly reduced impact on the sensitivity. And any observed asymmetry can be directly attributed to specific dimension-six interactions.

Depending on the specific process under consideration, a range of observables have been considered in the literature. In weak boson fusion, $pp\to H jj$, the most prominent of such observables is the `signed' azimuthal angle between the two jets~\cite{Plehn:2001nj}
\begin{equation}
\label{eq:phijj}
\Delta\phi_{jj} = \phi({j_1}) -\phi({j_2})~\text{for}~\eta(j_1) > \eta(j_2).
\end{equation}
Including transverse momentum information, this observable has been improved to an `optimal observable', considered in, e.g.~\cite{ATLAS:2016ifi, CMS:2019jdw, ATLAS:2020evk, CMS:2021nnc}. The (C)P properties of this observable can be exported to a range of other processes by constructing similar signed angular observables. Examples include the leptonic equivalent of this observable~\cite{Buckley:2015ctj, Englert:2019xhk, Goncalves:2021dcu},
\begin{equation}
\Delta\phi_{\ell\ell} = \phi({\ell_1}) -\phi({\ell_2})~\text{for}~\eta(\ell_1) > \eta(\ell_2),
\end{equation}
which is relevant for Higgs CP measurements using $ZH$ production at both electron-positron and proton-proton colliders.

Moving to cleaner, fully reconstructible final states such as $H\to ZZ^\ast\to 4\ell$, a wide variety of angular observables can be constructed~\cite{Buszello:2002uu, Choi:2002jk, Buszello:2004be} to measure and constrain coupling properties of the gauge-Higgs interactions. Specifically, CP-violating effects in $H\to 4\ell$ can be investigated using the variable $\Phi_{4\ell{}}$, as defined in Refs. \cite{Bolognesi:2012mm,Gritsan:2016hjl}
\begin{subequations}
\label{eq:cpodd}
\begin{equation}
    \Phi_{4\ell{}} = \frac{{\bf q}_1 \cdot ( \hat{\bf n}_1 \times \hat{\bf n}_2)}{| {\bf q}_1 \cdot \left( \hat{\bf n}_1 \times \hat{\bf n}_2\right)|} \; {\arccos}({\bf \hat{\bf n}_1 \cdot \hat{\bf n}_2}),
\end{equation}
where the normal vectors to the decay planes are given by  
\begin{equation}
    {\hat{\bf n}_1} = \frac{{\bf q}_{11} \times {\bf q}_{12}}{| {\bf q}_{11} \times {\bf q}_{12}|}\quad{\text{and}}\quad\hat{\bf n}_2 = \frac{{\bf q}_{21} \times {\bf q}_{22}}{|{\bf q}_{21} \times {\bf q}_{22}|}.
\end{equation}
\end{subequations}
Here, each ${\bf q}_{\alpha\beta}$ represents the three-momentum of the lepton or antilepton $\beta$ originating from the decay $Z_\alpha \to \ell^- {\ell}^+$, while ${\bf q}_\alpha = {\bf q}_{\alpha1} + {\bf q}_{\alpha2}$ denotes the total three-momentum of the corresponding $Z_\alpha$. All momenta are evaluated in the Higgs boson centre-of-mass frame.  Notably, the variable $\Phi_{4\ell{}}$ corresponds to the difference in polar angles between the same-charge leptons in their respective $Z$-boson rest frames, provided the reference axes are aligned. 

More recently, machine learning has been employed to optimise beyond traditional collider observables, significantly improving the sensitivity of CP-violation searches~\cite{Brehmer:2017lrt, Gritsan:2020pib, Bhardwaj:2021ujv, Hall:2022bme,ATLAS:2025edf}. Depending on the operator structures considered, these techniques can dramatically enhance sensitivity. For instance, at the LHC, incorporating additional characteristic kinematic information into an optimised asymmetry observable has been shown to improve sensitivity by up to a factor of ten. This is achieved (see also \cite{Bhardwaj:2021ujv}) by categorising the interference contribution into positively-weighted and negatively-weighted events depending on the sign of the differential event probability. This serves as two target groups for binary classification tasks. To further enhance the separation between interference and SM contributions, the SM prediction can be incorporated into the training process in a multiclass approach.\footnote{The scalability to many competing interactions has been demonstrated in~\cite{Atkinson:2021jnj} for non CP-violating interactions.} This serves the purpose of further optimising observables when traditional kinematics such as $\Delta \phi_{jj}$ are near optimal~\cite{Bhardwaj:2021ujv}, leading to reduced statistical information in the associated asymmetry. From this, CP-sensitive observables can be defined on an event-by-event basis using the trained models' output
\begin{equation}
\label{eq:mlcp}
O_{\NN} = P_+ - P_-
\end{equation}
where $P_+$ and $P_-$ represent the probabilities that an event belongs to the positively-weighted or negatively-weighted category, respectively. In a binary classification setup, these probabilities satisfy $P_+ + P_- = 1$. For a multiclass model, the sum extends to the SM probability, such that $P_+ + P_- + P_{\SM} = 1$; $P_{\SM}$ denotes the probability of an event being classified as an SM event.\footnote{$O_{\NN}$ can be constructed using neural networks (NNs) or boosted decision trees (BDTs). Both approaches have been investigated in this paper, with similar results obtained for each architecture. The NNs are trained using the MLPClassifier in Scikit HEP~\cite{Rodrigues:2020syo}, whereas the boosted decision trees were trained using CatBoost~\cite{DBLP:journals/corr/DorogushGGKPV17}. Hyperparameter optimisation is performed individually for each architecture. To mitigate biases from statistical fluctuations, a data augmentation procedure is applied during training. Each event is used twice: once with its original input variables and once with a CP transformation applied to these variables. For interference samples, the event weight is multiplied by $-1$ for CP-flipped events.}

For any of these differential observables, whether they relate to traditional collider observables Eqs.~\eqref{eq:phijj}-\eqref{eq:cpodd} or are obtained from ML-optimisation Eq.~\eqref{eq:mlcp}, it is possible to construct theoretically appealing asymmetries that will show robust performance as the dominant backgrounds and systematics are expected to be symmetric.

%%%%%%%%%%%%%%%%%%%%%%%%%%%%%%%%%%%%%%%%%%%%%%%%%%%%%
\section{Limit-setting procedure}
\label{sec:limits}
%%%%%%%%%%%%%%%%%%%%%%%%%%%%%%%%%%%%%%%%%%%%%%%%%%%%%
To determine the sensitivity, we construct confidence intervals for the process-dependent CP-odd observables using a binned likelihood function
\begin{equation}
\label{eq:llh}
 {\cal{L}}(\{c_i\}/\Lambda^2) =  \prod_{k} \exp\{-\lambda_k\} \frac{\lambda_k^{n_k}}{n_k!} .
\end{equation}
Here, $k$ labels the bins, with $n_k$ representing the expected number of events in bin $k$ under the SM-only hypothesis. The predicted number of events, $\lambda_k $, is derived from the SMEFT hypothesis for a given choice of the Wilson coefficients. The likelihood function in Eq.~\eqref{eq:llh} is converted into a confidence level using a profile-likelihood test statistic~\cite{Feldman:1997qc}. Through Wilks’ theorem~\cite{Wilks:1938dza}, this statistic follows a $\chi^2$ distribution with one degree of freedom, allowing us to extract the 95\% confidence level. The values of $\lambda_k $ and $n_k $ in Eq.~\eqref{eq:llh} are obtained directly from simulated samples, after applying normalisation factors to convert predicted cross sections into event yields. These normalisation factors are determined for each process to ensure that the SM prediction matches the expected number of SM events within the relevant fiducial regions of the corresponding analyses; details will be given for the relevant processes and colliders below.

Systematic uncertainties are not explicitly included in the likelihood function, as the constraints primarily arise from asymmetries in CP-odd observables. Since systematic uncertainties generally affect these distributions symmetrically, their impact is expected to be small~\cite{Hall:2022bme}. 

%%%%%%%%%%%%%%%%%%%%%%%%%%%%%%%%%%%%%%%%%%%%%%%%%%%%%
\section{$ZH$ production at $e^+ e^-$ colliders}
\label{sec:eezh}
%%%%%%%%%%%%%%%%%%%%%%%%%%%%%%%%%%%%%%%%%%%%%%%%%%%%%
The sensitivity of $ZH$ production at electron-positron colliders to CP-violating $HZZ$ interactions is studied for inclusive Higgs boson decays using samples of $e^+e^-\rightarrow \ell^+\ell^- H$. Simulated samples are produced with a version of \mg{} that does not account for initial state radiation (ISR) in $e^+e^-$ collisions. The cross section is, hence, corrected by a factor $ k_\textrm{isr}=\sigma_{ZH}^{\rm isr \,\, on}/\sigma_{ZH}^{\rm MG}$, where $\sigma_{ZH}^{\rm MG}$ is the \mg{} $e^+e^-\rightarrow ZH$ cross section and $\sigma_{ZH}^{\rm isr \,\, on}$ is the $e^+e^-\rightarrow ZH$ cross section with ISR effects included~\cite{Azzi:2012yn}. The dominant backgrounds from $ZZ$ production ($e^+e^-\rightarrow \ell^+\ell^- Z$) and $WW$ production ($e^+e^-\rightarrow W^+ W^-$) are also simulated and normalised using $k_\textrm{isr}$.

%%%%%%%%%%%%%%%%%%%%%%%%%%%%%%%%%%%%%%%%%%%%%%%%%%%%%
\subsection{Inclusive Higgs boson decays at FCC-ee}
\label{sec:eezh_fcc}
%%%%%%%%%%%%%%%%%%%%%%%%%%%%%%%%%%%%%%%%%%%%%%%%%%%%%
Events are required to satisfy selection criteria consistent with a $e^+e^-\rightarrow ZH$ final state, following the approach taken in Ref.~\cite{Azzi:2012yn}. Two same-flavour leptons are required, each with $p_{\rm T}>10$~GeV. The momentum of the dilepton system is required to satisfy $p_{\rm T}^{\ell\ell}>15$~GeV and $p_{\rm z}^{\ell\ell}<70$~GeV. The angle between the two leptons has to satisfy ${\arccos} ({\vec{n}}_{\ell_{1}} \cdot {\vec{n}}_{\ell_{2}}) > 100^{\circ}$, where ${\vec{n}}_{\ell_{i}}$ is the normalised three-momentum of lepton $i$. The angle between the plane containing the two leptons and the $z$-axis is required to satisfy ${\arccos} ([{\vec{n}}_{\ell_{1}} \times {\vec{n}}_{\ell_{2}})\cdot {\vec{\hat{k}}}]) > 10^{\circ}$. The invariant mass of the dilepton system is required to satisfy $|m_{\ell\ell}-m_Z|<15$~GeV. The mass recoiling against the dilepton system can be calculated from  
$m_{\rm recoil}^2= s + m_{\ell\ell}^2 + 2\sqrt{s}(E_{\ell_1} + E_{\ell_2})$, where $E_{\ell_i}$ is the energy of lepton $i$, and is required to be consistent with a Higgs boson candidate, i.e. $|m_{\rm recoil}-m_H|<5$~GeV. Following this selection, 77,300 signal events are predicted for an integrated luminosity of 10.8~ab$^{-1}$ at $\sqrt{s}=240$~GeV. The background yields are 3,470 and 26,100 for $e^+e^-\rightarrow ZZ, W^+ W^-$, respectively.

\begin{figure}[p]
  \centering
  \subfigure[\label{fig:eezh_distributionsa}]{\includegraphics[width=0.46\textwidth]{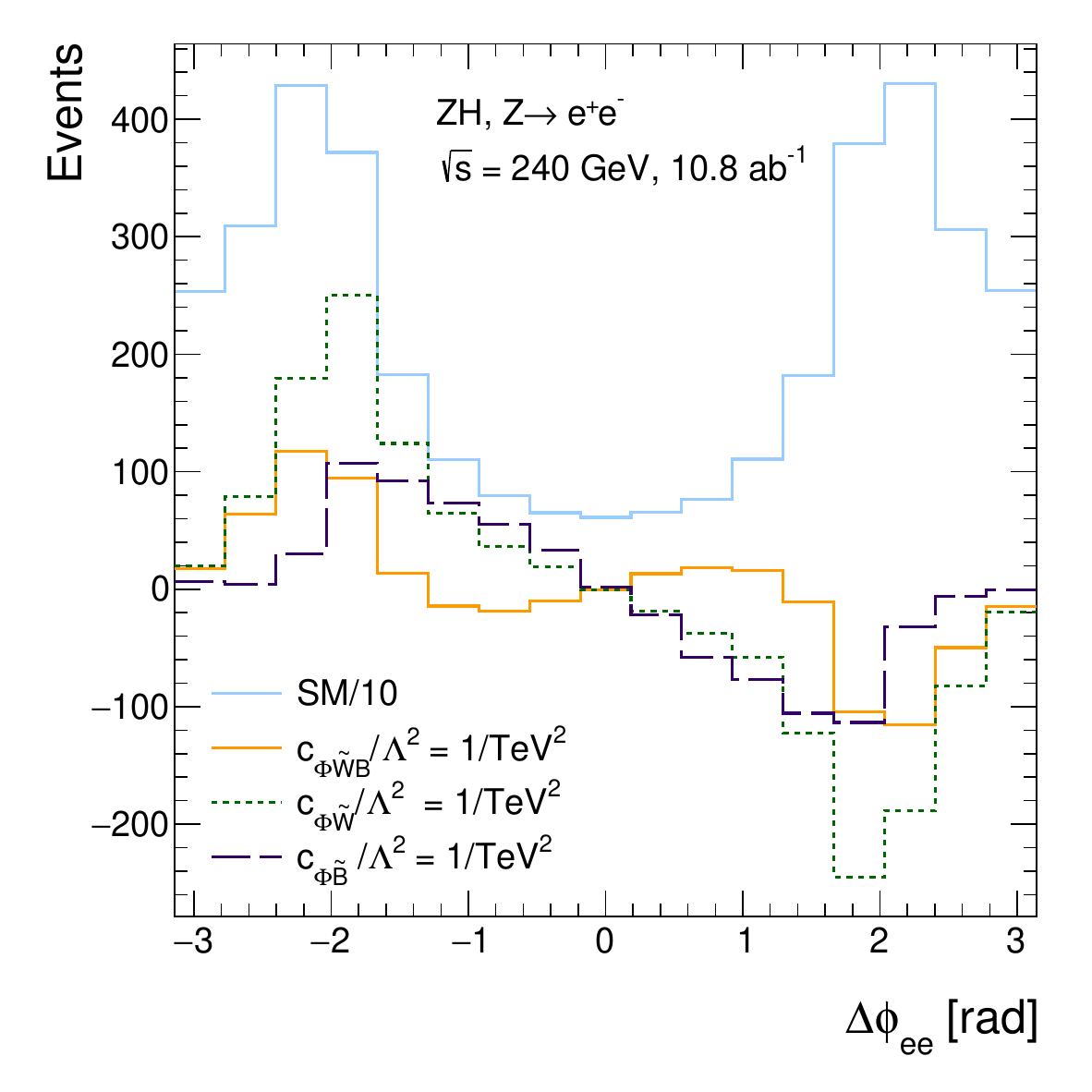}}
  \subfigure[\label{fig:eezh_distributionsb}]{\includegraphics[width=0.46\textwidth]{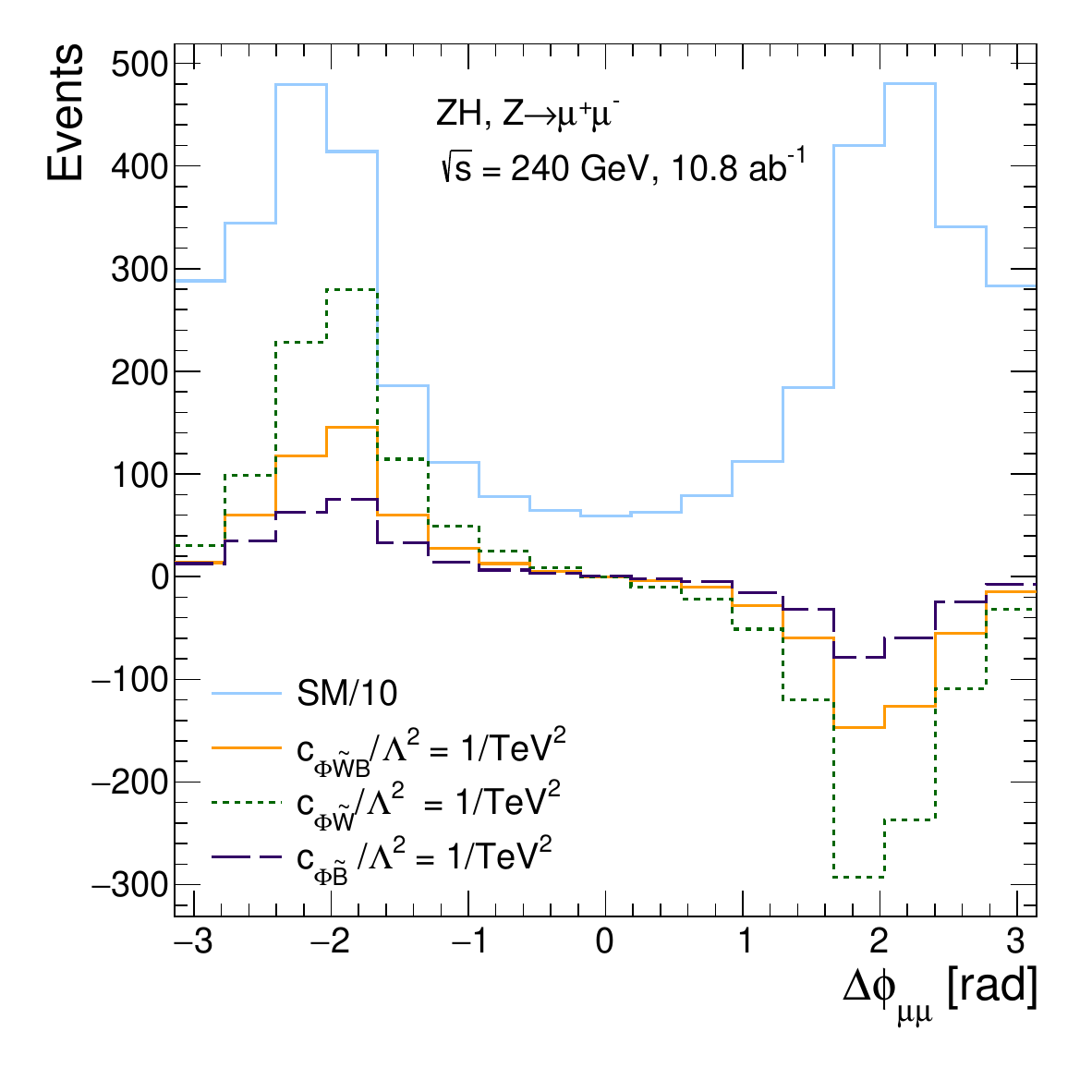}} \\
  \subfigure[\label{fig:eezh_distributionsc}]{\includegraphics[width=0.46\textwidth]{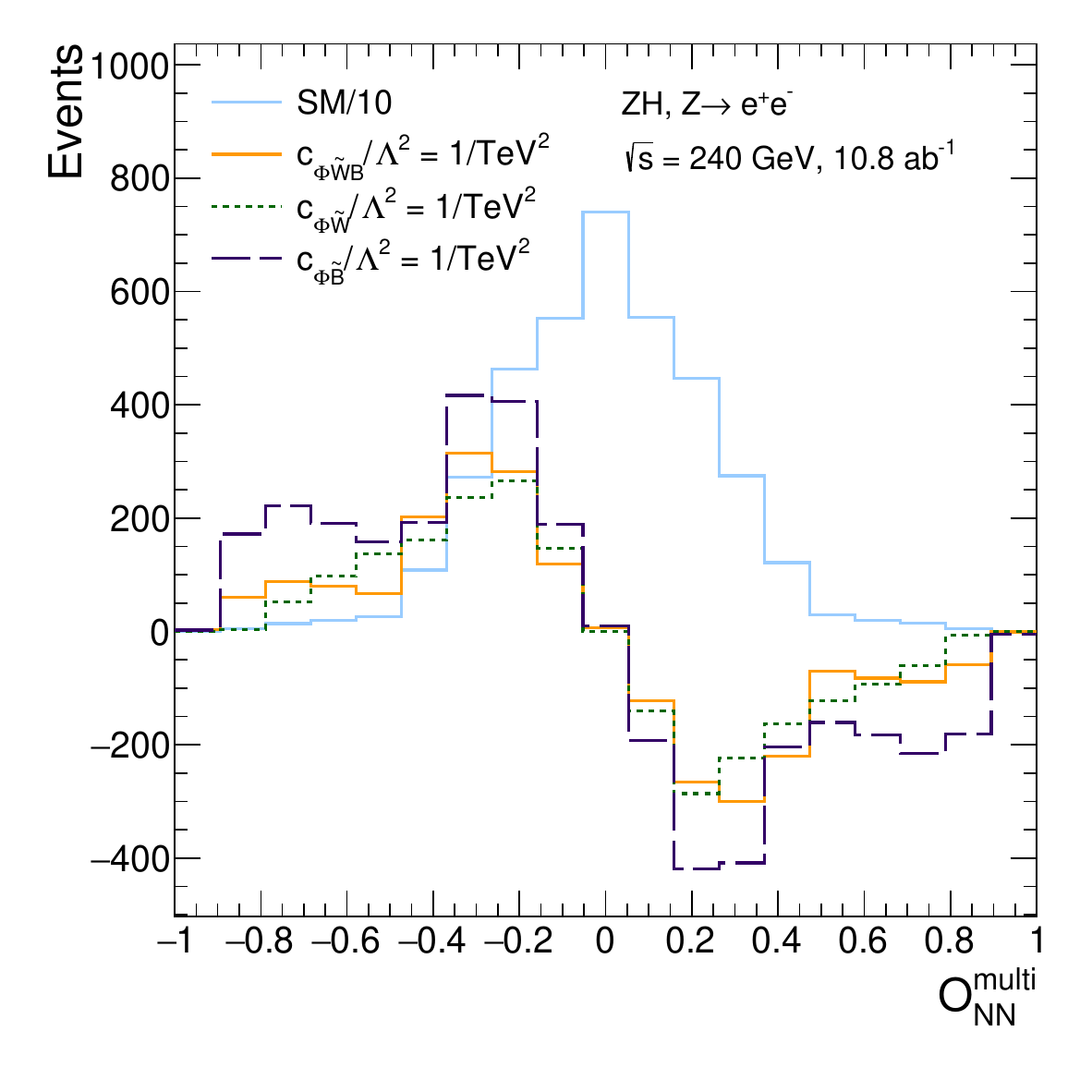}} 
  \hspace{0.01\textwidth}
  \subfigure[\label{fig:eezh_distributionsd}]{\raisebox{5mm}{\includegraphics[width=0.42\textwidth]{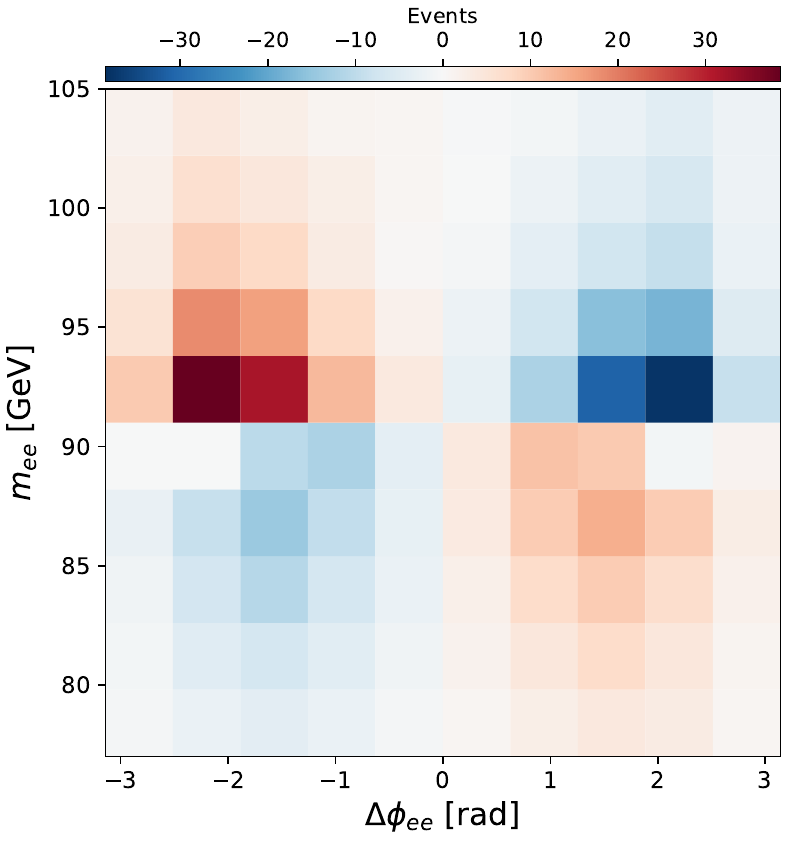}}}
  \caption{Expected event yields for $e^+e^-\rightarrow ZH$ production at FCC-ee as a function of $\Delta\phi_{\ell\ell}$ in (a) the electron decay channel of the $Z$ boson and (b) the muon decay channel of the $Z$ boson. (c) Expected event yield as a function of the ML-based observable ($O_{\rm NN}^{\rm multi}$) in the electron channel. (d) Expected event yields as a function of $m_{\ell\ell}$ and $\Delta\phi_{\ell\ell}$, for the interference contributions induced by the \OHWtilB{} operators in the electron channel. 
  \label{fig:eezh_distributions}}
\end{figure}

Two types of CP-odd observables are found to be sensitive to CP violation in the $HZZ$ coupling. The first is the signed azimuthal angle between the two leptons that originate from the $Z$-boson decay, $\Delta \phi_{\ell\ell}$. Secondly, there are the aforementioned ML-based observables constructed using either binary classifiers or multiclass models trained on the relevant interference samples. Figures~\ref{fig:eezh_distributionsa} and~\ref{fig:eezh_distributionsb} show the expected event yields as a function of $\Delta\phi_{\ell\ell}$ in the muon and electron channels, respectively, for both the SM prediction and the interference contributions produced by each dimension-six operator. There are clear differences in the shapes of the interference spectra in the two decay channels. Figure~\ref{fig:eezh_distributionsc} shows the $O_{\rm NN}^{\rm multi}$ observables for the \OHWtilB{} operator in the electron channel, which demonstrates much improved sensitivity compared to $\Delta\phi_{\ell\ell}$ alone.  An investigation into the origin of this sensitivity using permutation feature importance techniques exposes a sign-flip in the interference contribution for dilepton invariant masses around the $m_Z$ pole. This is shown in Fig.~\ref{fig:eezh_distributionsd}. The sign-flip in the interference around the $m_Z$ pole is only observed in the electron channel and not in the muon channel, which implies that the effect arises from vector-boson-fusion contributions to the scattering amplitude. 

%%%%%%%%%%%%%%%%%%%%%%%%%%%%%%%%%%%%%%%%%%%%%%%%%%%%%
%%%%%%%%%%%%%%%%%%%%%%%%%%%%%%%%%%%%%%%%%%%%%%%%%%%%%

The 95\% confidence intervals on the \cHBtil$/\Lambda^2$, \cHWtil$/\Lambda^2$, and \cHWtilB$/\Lambda^2$ Wilson coefficients are presented in Tab.~\ref{tab:eezh_limits}. The profile likelihood is performed after splitting the events into the electron channel and the muon channel. In the case of the \cHBtil$/\Lambda^2$ and \cHWtilB$/\Lambda^2$ Wilson coefficients, the constraints are improved by a factor of 4-5 when using a fit to the $O_{\rm NN}$ distribution. The improvements in sensitivity when using $O_{\rm NN}$ are driven by the electron channel, which exploits the sign-flip in the interference pattern about the $m_Z$ pole. The multiclass models do not add noticeable sensitivity compared to the binary classifiers; this is due to the limited kinematic range when running at the $ZH$ threshold at electron-positron colliders.

%%%%%%%%%%%%%%%%%%%%%%%%%%%%%%%%%%%%%%%%%%%%%%%%%%%%%
\begin{table}[t]
\centering
\begin{tabular}{cccc}
\toprule 
 \multirow{2}{*}{Observable} & \multicolumn{3}{c}{95\% confidence intervals} [TeV$^{-2}$] \\ \cmidrule{2-4}
 & \cHWtil$/\Lambda^2$ & \cHBtil$/\Lambda^2$ & \cHWtilB$/\Lambda^2$  \\ \midrule
$\Delta\phi_{ee}$ & [-0.25,0.25] & [-0.36,0.36] & [-0.57,0.57] \\ 
$\Delta\phi_{\mu\mu}$ & [-0.24,0.24] & [-0.89,0.89] & [-0.47,0.47] \\
$\Delta\phi_{\ell\ell}$ & [-0.17,0.17] & [-0.33,0.33] & [-0.36,0.36] \\
 \midrule
  $O_{\rm NN}^{\rm binary}$ & [-0.13,0.13] & [-0.08,0.08] & [-0.12,0.12] \\
  $O_{\rm NN}^{\rm multi}$ & [-0.13,0.13] & [-0.07,0.07] & [-0.12,0.12] \\
\bottomrule
\end{tabular}
\caption{95\% confidence intervals on the \cHWtil$/\Lambda^2$, \cHBtil$/\Lambda^2$, and \cHWtilB$/\Lambda^2$ Wilson coefficients using $e^+e^-\rightarrow ZH$ production at FCC-ee. Constraints are shown separately for the electron channel, the muon channel, and their combination. The $O_{\rm NN}^{\rm binary}$ observables are constructed using a binary classifier. The $O_{\rm NN}^{\rm multi}$ observables are constructed using a multiclass model that discriminates between constructive interference, destructive interference, and the SM.}
\label{tab:eezh_limits}
\end{table}
%%%%%%%%%%%%%%%%%%%%%%%%%%%%%%%%%%%%%%%%%%%%%%%%%%%%%

%%%%%%%%%%%%%%%%%%%%%%%%%%%%%%%%%%%%%%%%%%%%%%%%%%%%%
\subsection{Extension to the $H\rightarrow b\bar{b}$ decay channel}
%%%%%%%%%%%%%%%%%%%%%%%%%%%%%%%%%%%%%%%%%%%%%%%%%%%%%
The analysis presented in Sec.~\ref{sec:eezh_fcc} can be extended to the $H\rightarrow b\bar{b}$ decay channel by selecting the subset of events that have at least two $b$-tagged jets with $p_{\rm T}>20$~GeV. Dedicated samples of inclusive $e^+e^-\rightarrow \ell^+\ell^- b \bar{b}$ events are produced, which contain both the signal ($ZH$) and background ($ZZ$) processes. Following this selection, 12,900 $\ell^+\ell^- b\bar{b}$ events are predicted for an integrated luminosity of 10.8~ab$^{-1}$ at 240 GeV.\footnote{The event yields obtained after requiring two $b$-jets have been studied for both the purpose-generated $e^+e^-\rightarrow \ell^+\ell^- b \bar{b}$ sample and the combined $e^+e^-\rightarrow \ell^+\ell^-H$ and $e^+e^-\rightarrow \ell^+\ell^-Z$ samples used in Sec.~\ref{sec:eezh_fcc}. For the SM, the event yields are nearly 20\% larger when using the $e^+e^-\rightarrow \ell^+\ell^- b \bar{b}$ sample, across all distributions. For the interference samples, the effect is larger and the amplitude of the interference spectrum as a function of $\Delta\phi_{\ell\ell}$ is increased by nearly a factor of two. These effects arise from interferences between signal diagrams (containing a Higgs boson) and background diagrams (no Higgs boson). We note that (i) such large interference effects between Higgs and non-Higgs processes have been largely overlooked in previous electron-proton collider studies and (ii) that the constraints achieved in Sec.~\ref{sec:eezh_fcc} are likely conservative.}

Observables constructed from the $b$-jets are found to be insensitive to the CP-violating effects in the $HZZ$ coupling, whereas the interference structures in the $\Delta\phi_{\ell\ell}$ distribution are found to be similar to those observed for inclusive Higgs boson decays. For the ML-based approach, a detailed investigation of the binary classifier showed that the performance is almost entirely driven by the information in the $Z\rightarrow \ell^+\ell^-$ decay.\footnote{The observation that the Higgs boson decay products carry no relevant information on CP-violating effects in the $HZZ$ coupling was cross-checked in the $H\rightarrow\tau\tau$ channel. Similar results were obtained.} 

%%%%%%%%%%%%%%%%%%%%%%%%%%%%%%%%%%%%%%%%%%%%%%%%%%%%%
\begin{figure}[t]
  \centering
  \subfigure[]{\includegraphics[width=0.46\textwidth]{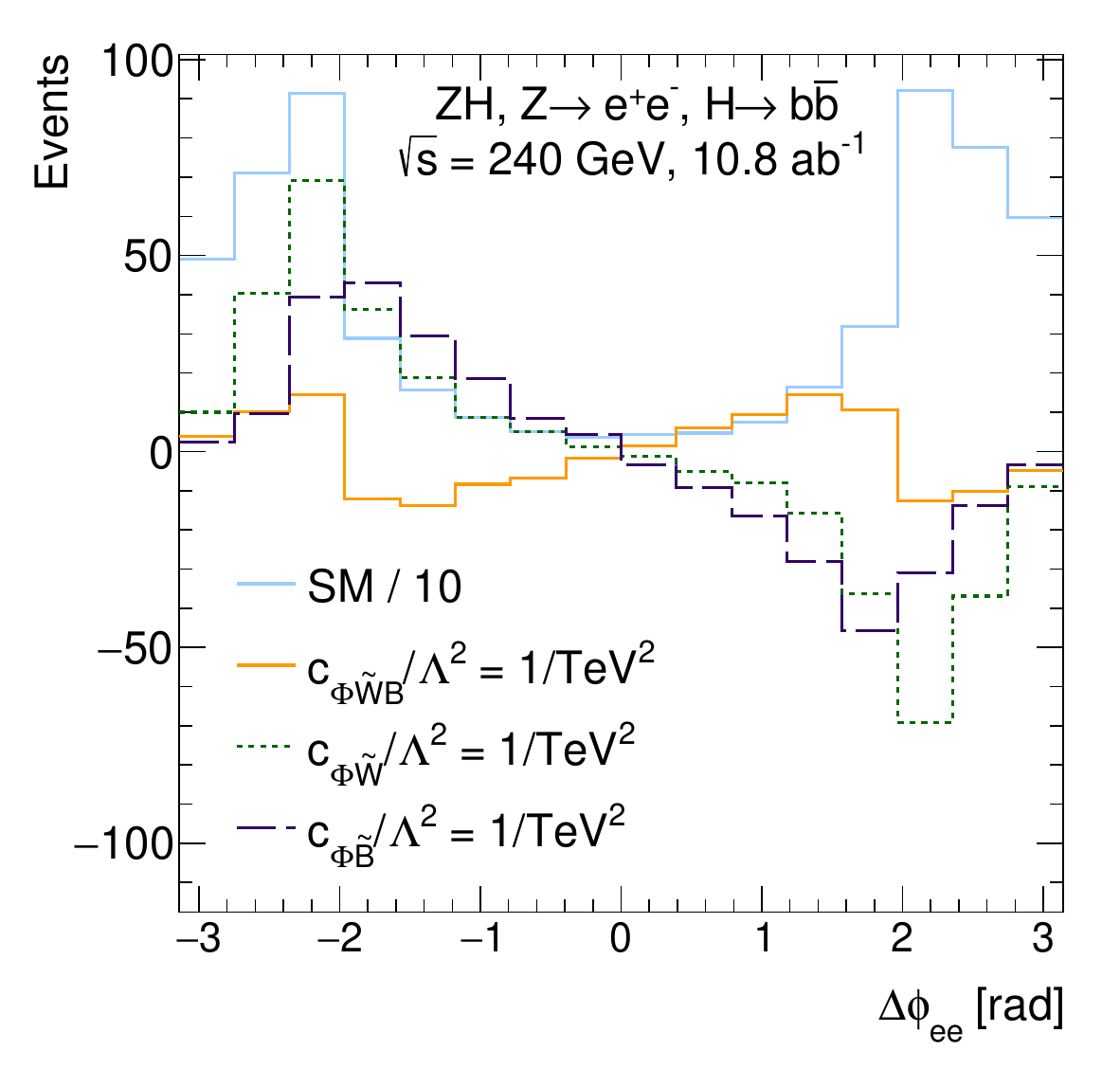}}
  \subfigure[]{\includegraphics[width=0.46\textwidth]{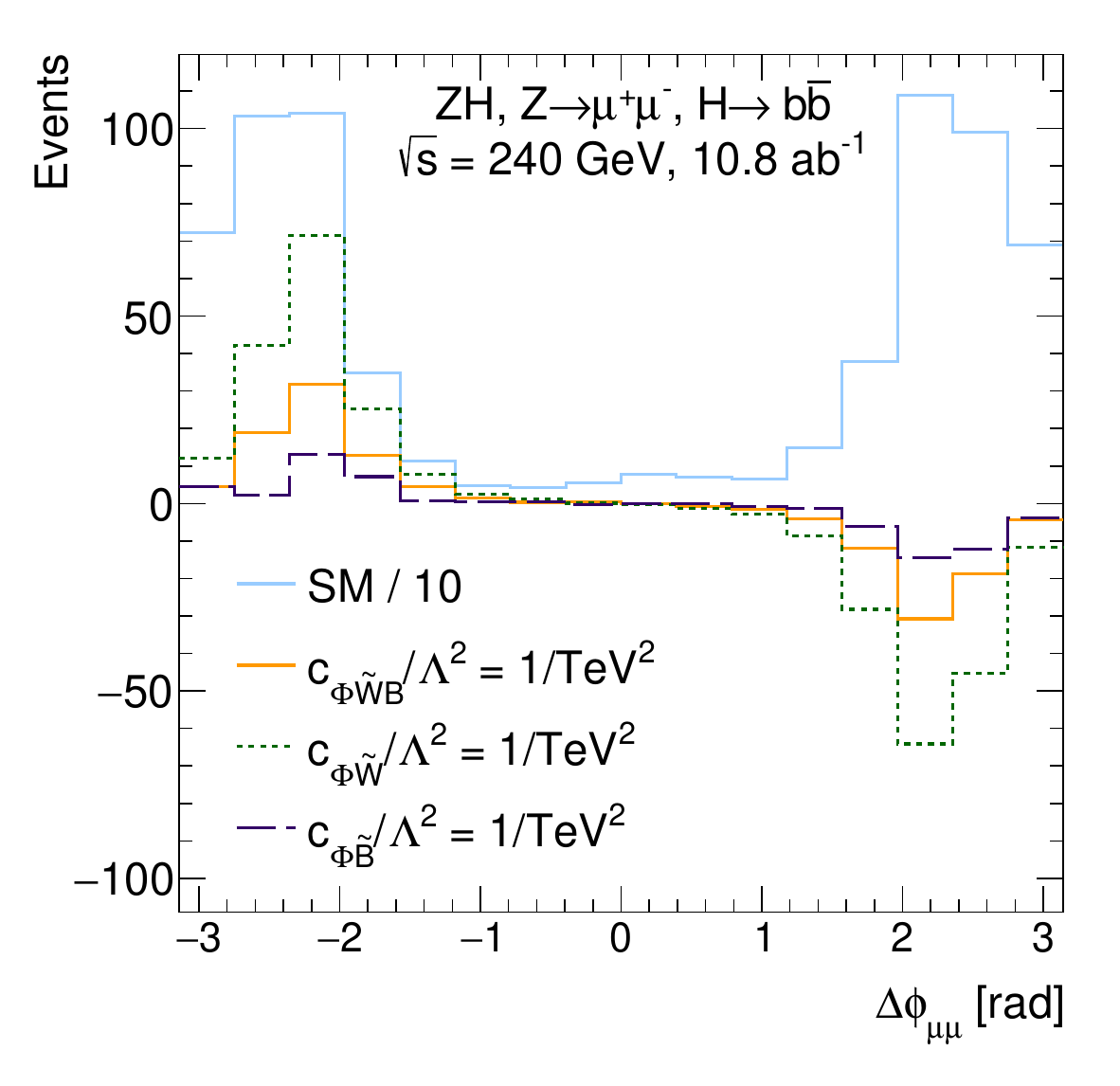}} \\
  \caption{Expected event yields for $e^+e^-\rightarrow ZH$ production in the $H\rightarrow b\bar{b}$ decay channel at FCC-ee. Distributions are shown as a function of $\Delta\phi_{\ell\ell}$ in (a) the electron decay channel of the $Z$ boson (a) and (b) the muon decay channel of the $Z$ boson.  
  \label{fig:eezh_hbb_distributions}}
\end{figure}
%%%%%%%%%%%%%%%%%%%%%%%%%%%%%%%%%%%%%%%%%%%%%%%%%%%%%

The 95\% confidence intervals on the \cHBtil$/\Lambda^2$, \cHWtil$/\Lambda^2$, and \cHWtilB$/\Lambda^2$ Wilson coefficients are presented in Tab.~\ref{tab:eezhbb_limits}. The results follow the same pattern as for inclusive Higgs boson decays, with ML-based observables outperforming simple angular observables. In the muon channel, the sensitivity in the $H\rightarrow b\bar{b}$ channel is approximately a factor of two worse than that obtained for inclusive Higgs decays. This can be understood from the reduced event yields caused by the Higgs branching ratio, the inefficiency of reconstructing two $b$-jets, and the selection inefficiency when requiring two jets with $p_{\rm T}>20$~GeV. In the electron channel, however, the sensitivity in the $H\rightarrow b\bar{b}$ channel is much closer to that obtained for inclusive Higgs boson decays. This feature has been studied in detail and arises from a different selection efficiency when requiring two $b$-jets for SM events and interference events. In particular, the differences in the selection efficiency between the SM and interference samples are topology-dependent and the ratio of interference-to-SM increases for the \OHBtil{} and \OHWtilB{} operators in the region $|\Delta\phi_{\ell\ell}| \leq 1.5$ in the $H\rightarrow b\bar{b}$ analysis, as shown in Fig.~\ref{fig:eezh_hbb_distributions}.

%%%%%%%%%%%%%%%%%%%%%%%%%%%%%%%%%%%%%%%%%%%%%%%%%%%%%
\begin{table}[t]
\centering
\begin{tabular}{cccc}
\toprule 
 \multirow{2}{*}{Observable} & \multicolumn{3}{c}{95\% confidence intervals} [TeV$^{-2}$] \\ \cmidrule{2-4}
 & \cHWtil$/\Lambda^2$ & \cHBtil$/\Lambda^2$ & \cHWtilB$/\Lambda^2$  \\ \midrule
$\Delta\phi_{ee}$ & [-0.35,0.35] & [-0.31,0.31] & [-0.67,0.67] \\
$\Delta\phi_{\mu\mu}$ & [-0.46,0.46] & [-2.2,2.2] & [-1.0,1.0] \\
$\Delta\phi_{\ell\ell}$ & [-0.28,0.28] & [-0.31,0.31] & [-0.56,0.56] \\
 \midrule
  $O_{\rm NN}^{\rm binary}$ & [-0.23,0.23] & [-0.08,0.08] & [-0.14,0.14] \\
\bottomrule
\end{tabular}
\caption{95\% confidence intervals on the \cHWtil$/\Lambda^2$, \cHBtil$/\Lambda^2$, and \cHWtilB$/\Lambda^2$ Wilson coefficients at the FCC-ee, obtained using $e^+e^-\rightarrow ZH$ production and in the $H\rightarrow b\bar{b}$ decay channel. Constraints are shown separately for the electron channel, the muon channel, and their combination. The $O_{\rm NN}^{\rm binary}$ observables are constructed using a binary classifier.}
\label{tab:eezhbb_limits}
\end{table}
%%%%%%%%%%%%%%%%%%%%%%%%%%%%%%%%%%%%%%%%%%%%%%%%%%%%%

%%%%%%%%%%%%%%%%%%%%%%%%%%%%%%%%%%%%%%%%%%%%%%%%%%%%%
\subsection{Impact of beam polarisation at LCF}
\label{sec:eezh_fccbeam}
%%%%%%%%%%%%%%%%%%%%%%%%%%%%%%%%%%%%%%%%%%%%%%%%%%%%%
In this section, the FCC-ee analysis presented in Sec.~\ref{sec:eezh_fcc} is modified to account for beam polarisation and lower luminosity at the LCF. Dedicated signal and background samples are produced at $\sqrt{s}=250$~GeV for each of the four initial-state polarisation configurations (discussed in Sec.~\ref{sec:setup}), which are then summed to produce the expected mix of polarisations expected at the LCF. The same event selection criteria are applied as in Sec.~\ref{sec:eezh_fcc}. Following this selection, 25,300 signal events are predicted for an integrated luminosity of 3~ab$^{-1}$ at $\sqrt{s}=250$~GeV. The corresponding background yields are 1,110 and 7,960 for $e^+e^-\rightarrow ZZ$ and $e^+e^-\rightarrow W^+ W^-$, respectively.

%%%%%%%%%%%%%%%%%%%%%%%%%%%%%%%%%%%%%%%%%%%%%%%%%%%%%
\begin{figure}[p]
  \centering
 \subfigure[]{\includegraphics[width=0.47\textwidth]{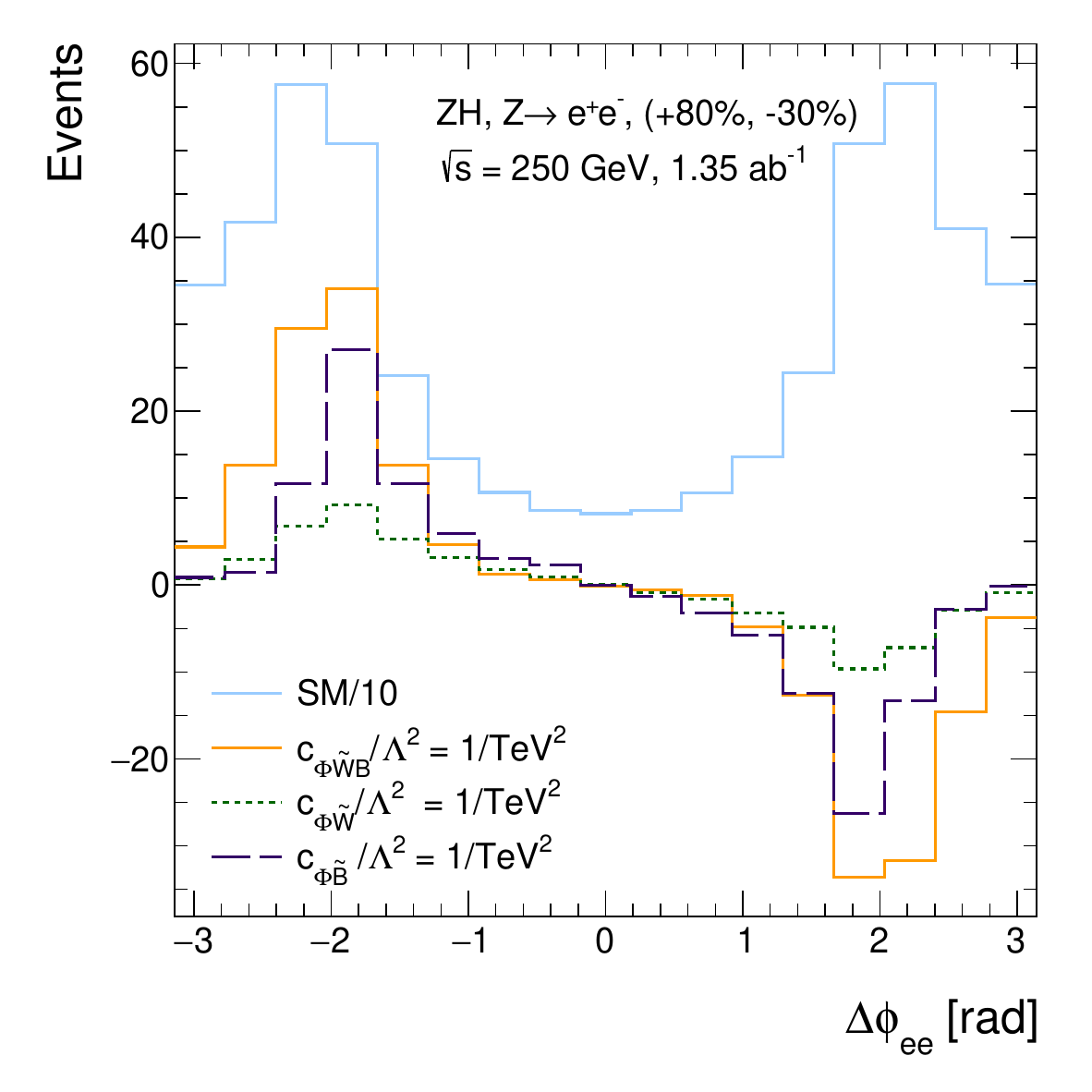}} \quad
  \subfigure[]{\includegraphics[width=0.47\textwidth]{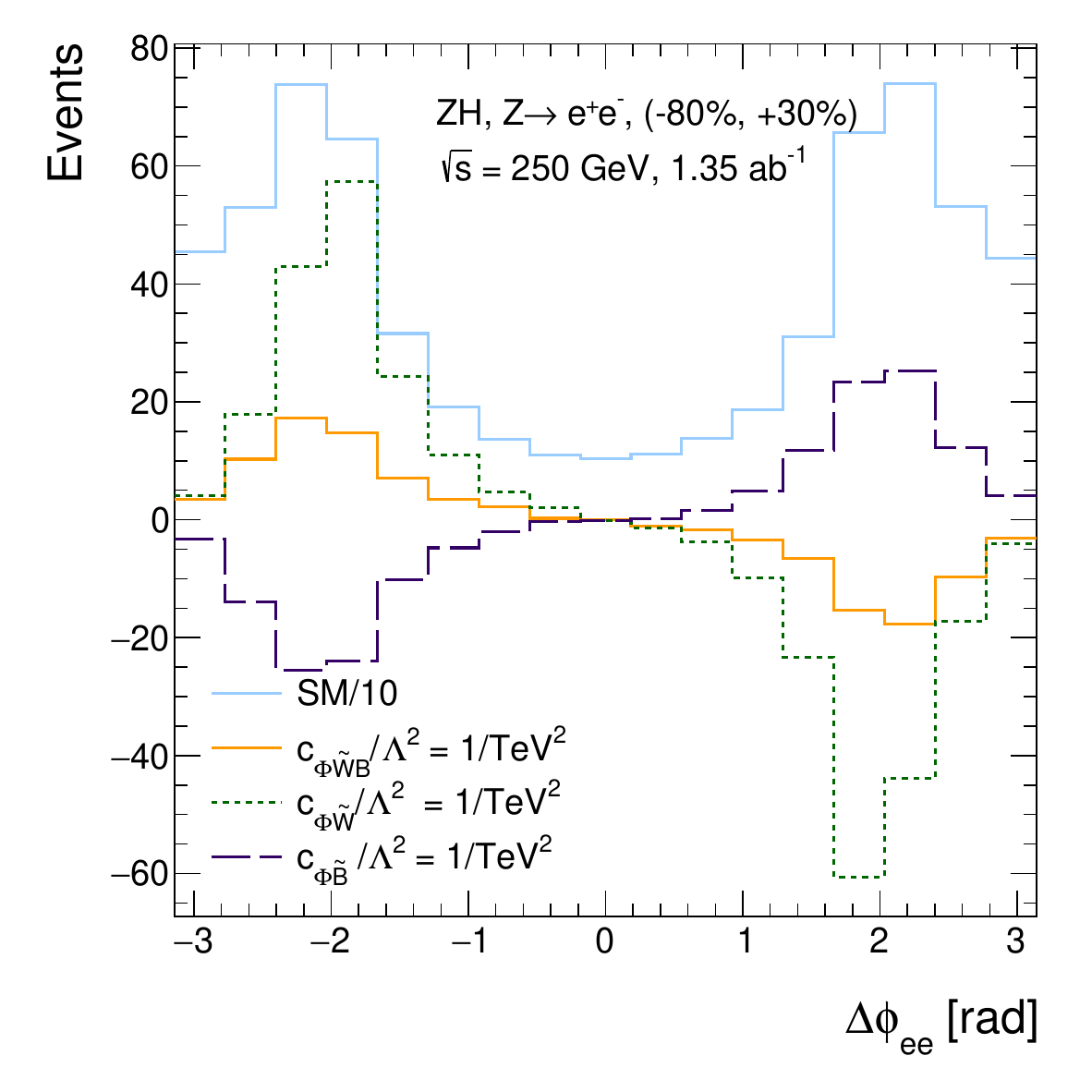}} \\
  \subfigure[]{\includegraphics[width=0.47\textwidth]{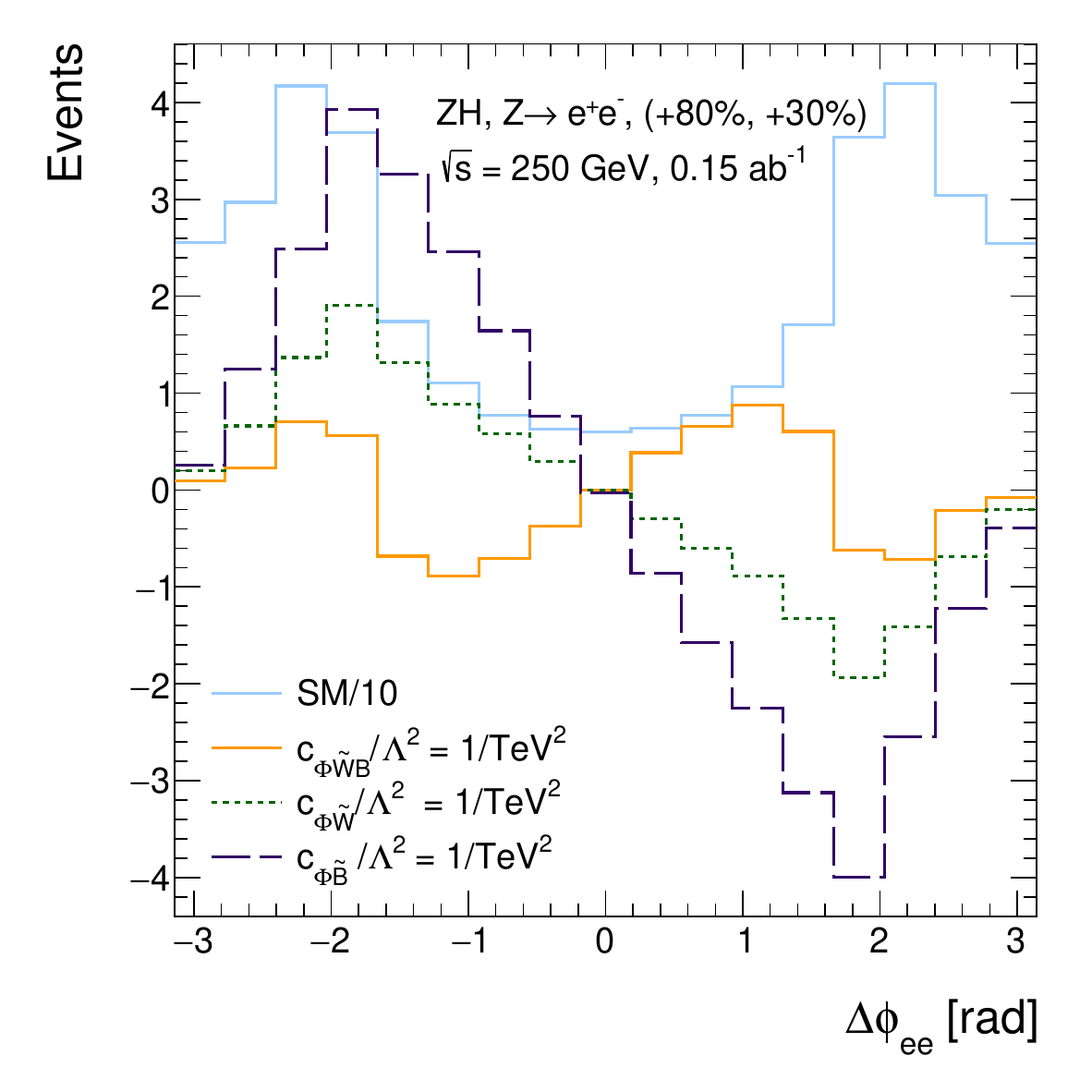}} \quad
  \subfigure[]{\includegraphics[width=0.47\textwidth]{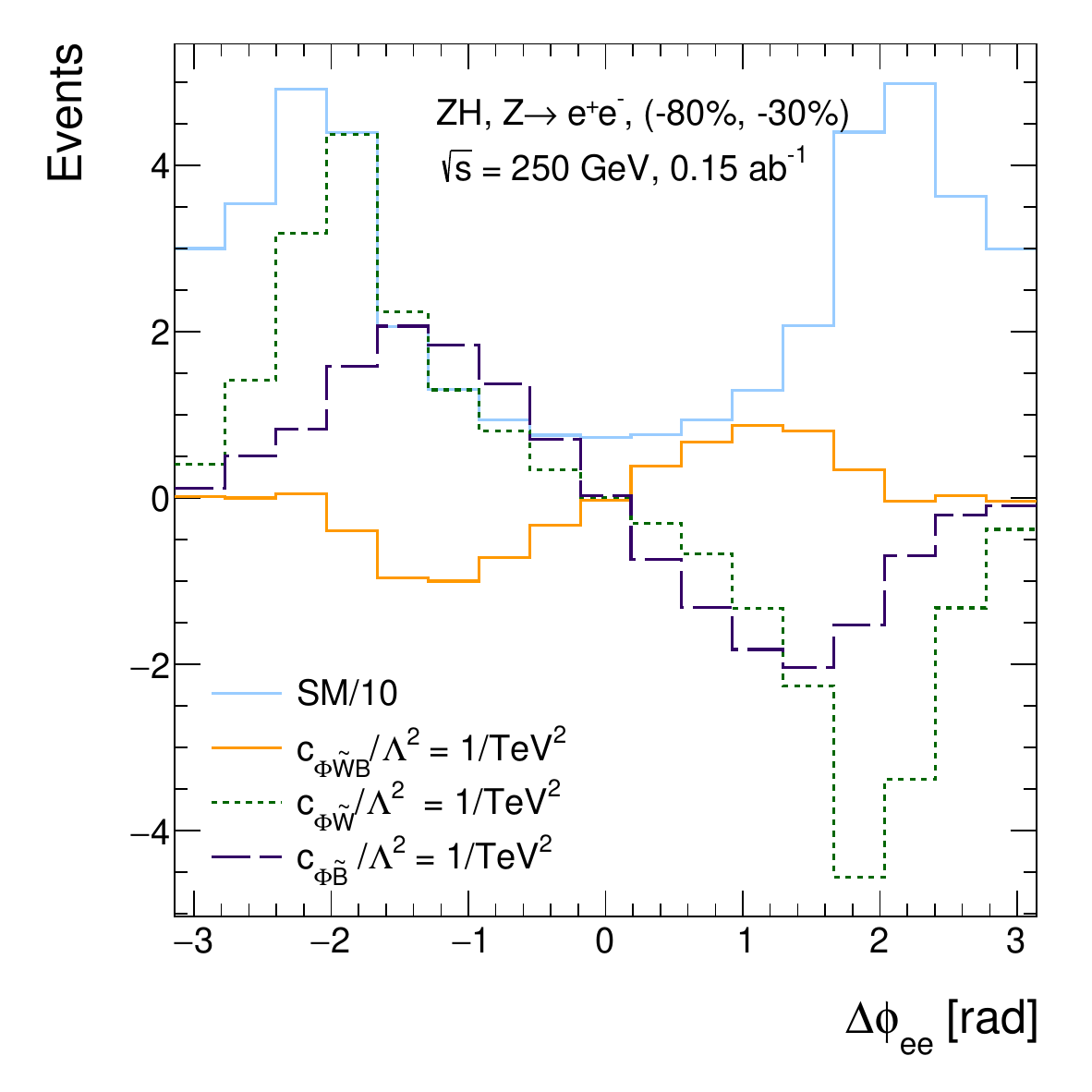}}
  \caption{Expected event yields for $e^+e^-\rightarrow ZH$ production at LCF with inclusive Higgs boson decays and in the electron decay channel of the $Z$ boson. Distributions are shown as a function of $\Delta\phi_{\ell\ell}$ for different initial-state beam polarisations, namely (a) $ \{ e^-_\textrm{pol}, e^+_\textrm{pol}\} = \{+80\%,-30\%\}$ (b) $ \{ e^-_\textrm{pol}, e^+_\textrm{pol}\} = \{-80\%,+30\%\}$, (c) $ \{ e^-_\textrm{pol}, e^+_\textrm{pol}\} = \{+80\%,+30\%\}$, and (d) $ \{ e^-_\textrm{pol}, e^+_\textrm{pol}\} = \{-80\%,-30\%\}$.
  \label{fig:eezh_lcf_distributions}}
\end{figure}
%%%%%%%%%%%%%%%%%%%%%%%%%%%%%%%%%%%%%%%%%%%%%%%%%%%%%

Figure~\ref{fig:eezh_lcf_distributions} shows the expected event yields as a function of $\Delta\phi_{\ell\ell}$ in the electron channel, for each of the initial state beam polarisation configurations. The interference structure induced by each operator is heavily dependent on the beam polarisation. For example, the interference induced by the \OHBtil{} operator flips sign between the $\{+80\%,-30\%\}$ and $\{-80\%,+30\%\}$ configurations. Furthermore, with respect to the SM, the magnitude of the interference contribution induced by this operator is much larger for the $\{+80\%,+30\%\}$ and $\{-80\%,-30\%\}$ configurations than for the $\{+80\%,-30\%\}$ and $\{-80\%,+30\%\}$ configurations. To exploit the sensitivity to different beam polarisations when constructing the ML-based observables, dedicated classifiers are trained for each initial-state polarisation configuration, and the $O_{\textrm NN}$ distributions are then combined in the likelihood when setting limits on the Wilson coefficients.

The 95\% confidence intervals on the \cHBtil$/\Lambda^2$, \cHWtil$/\Lambda^2$, and \cHWtilB$/\Lambda^2$ Wilson coefficients are presented in Tab.~\ref{tab:eezh_lcf_limits}. The results are shown for both polarised and unpolarised beams. The beam polarisation has a significant effect, improving the limits by a factor of 1.2-1.8 compared to the unpolarised case, depending on the distribution used in the limit setting. The improvement in sensitivity when exploiting beam polarisation at LCF partially mitigates the impact of smaller event yields caused by the lower luminosity at LCF (compared to FCC-ee).

%%%%%%%%%%%%%%%%%%%%%%%%%%%%%%%%%%%%%%%%%%%%%%%%%%%%%
\begin{table}[t]
\centering
\begin{tabular}{cccc}
\toprule 
 \multirow{2}{*}{Observable} & \multicolumn{3}{c}{95\% confidence intervals} [TeV$^{-2}$] \\ \cmidrule{2-4}
 & \cHWtil$/\Lambda^2$ & \cHBtil$/\Lambda^2$ & \cHWtilB$/\Lambda^2$  \\ \midrule
$\Delta\phi_{\ell\ell}$ (pol) & [-0.28,0.28] & [-0.35,0.35] & [-0.38,0.38] \\
$\Delta\phi_{\ell\ell}$ (unpol)  & [-0.33,0.33] & [-0.63,0.63] & [-0.69,0.69] \\
 \midrule
  $O_{\rm NN}^{\rm binary}$ (pol) & [-0.21,0.21] & [-0.11,0.11] & [-0.19,0.19]\\
  $O_{\rm NN}^{\rm binary}$ (unpol) & [-0.25,0.25] & [-0.14,0.14] & [-0.23,0.23] \\
\bottomrule
\end{tabular}
\caption{95\% confidence intervals on the \cHWtil$/\Lambda^2$, \cHBtil$/\Lambda^2$, and \cHWtilB$/\Lambda^2$ Wilson coefficients using $e^+e^-\rightarrow ZH$ production at LCF with an integrated luminosity of 3~ab$^{-1}$. Results are shown for the nominal beam polarisation at LCF. The effect of using unpolarised beams (labelled `unpol') with the same integrated luminosity is also shown for comparison. The $O_{\rm NN}^{\rm binary}$ observables are constructed using a binary classifier.}
\label{tab:eezh_lcf_limits}
\end{table}
%%%%%%%%%%%%%%%%%%%%%%%%%%%%%%%%%%%%%%%%%%%%%%%%%%%%%

%%%%%%%%%%%%%%%%%%%%%%%%%%%%%%%%%%%%%%%%%%%%%%%%%%%%%
\section{$ZH$ production at $pp$ colliders}
\label{sec:ppzh}
%%%%%%%%%%%%%%%%%%%%%%%%%%%%%%%%%%%%%%%%%%%%%%%%%%%%%
The sensitivity of $ZH$ production at $pp$ colliders is studied in the dominant $H\rightarrow b\bar{b}$ decay channel. An inclusive sample of $pp\rightarrow \ell^+\ell^- b\bar{b}$ events is generated at order $\alpha_{\rm EW}^4$, which includes contributions from both signal ($ZH$ production) and diboson backgrounds ($ZZ$ production). The cross section for the SM and interference samples is then scaled by a factor $\sigma^{\rm NNLO}_{ZH}/\sigma^{\rm MG}_{ZH}$, where $\sigma^{\rm NNLO}_{ZH}$ and $\sigma^{\rm MG}_{ZH}$ are the cross sections for $ZH$ production calculated by the LHC Higgs cross section working group \cite{LHCHiggsCrossSectionWorkingGroup:2016ypw,Cepeda:2019klc} and by \mg{}, respectively. The SM sample is then scaled by an additional factor to account for missing loop-induced $gg\rightarrow ZH$ contributions~\cite{Cepeda:2019klc}. Finally, an additional sample of $\ell^+\ell^- b\bar{b}$ events is generated at order $\alpha_{\rm EW}^2 \alpha_{\rm QCD}^2$ to account for the dominant $Z+{\rm jets}$ background.

Events are required to satisfy selection criteria consistent with a recent ATLAS analysis of $ZH$ production in the $H\rightarrow b\bar{b}$ decay channel~\cite{ATLAS:2020fcp}. Two same-flavour leptons are required, each with $p_{\rm T}>7$~GeV and $|\eta|<2.47$ (electrons) or $|\eta|<2.7$ (muons). The dilepton system is required to satisfy $p_{\rm T}^{\ell\ell}>150$~GeV and $81<m_{\ell\ell}<101$~GeV. Jets are reconstructed and required to have $p_{\rm T}>20$~GeV ($|\eta|<2.5$) or $p_{\rm T}>35$~GeV ($2.5<|\eta|<4.5$). Exactly two $b$-tagged jets are required, and at least one of these is required to have $p_{\rm T}>45$~GeV. The invariant mass of the $b$-jet pair is required to satisfy $110 < m_{b \bar{b}}<150$~GeV.\footnote{The selection criteria are validated by comparing the measured event yields in the phase space of the relevant ATLAS analysis~\cite{ATLAS:2020fcp}, which are the same criteria as used in this article except for (i) omitting the $m_{b\bar{b}}$ cut and (ii) adding cuts on the $\Delta R$ between the two $b-{\rm jets}$. Good agreement is observed for the $ZH+ZZ$ signal sample, whereas the background from $Z+{\rm jets}$ is underestimated by a factor of 1.3. Note that the background sample is only accurate to leading order in perturbative QCD, and the agreement in yields would likely be improved if the cross section were corrected for missing higher-order effects. The $Z+{\rm jets}$ background sample is therefore scaled by the factor of 1.3 in the studies at all colliders, in line with the expected QCD corrections~\cite{Boughezal:2015ded}.} Following this event selection, the signal yield for $ZH+ZZ$ production is predicted to be 133 events at LHC and 207,000 at FCC-hh. The background from $Z+{\rm jets}$ is estimated to be 3,300 at LHC and 13,900,000 at FCC-hh.

%%%%%%%%%%%%%%%%%%%%%%%%%%%%%%%%%%%%%%%%%%%%%%%%%%%%%
\begin{figure}[!t]
  \centering
  \subfigure[]{\includegraphics[width=0.49\textwidth]{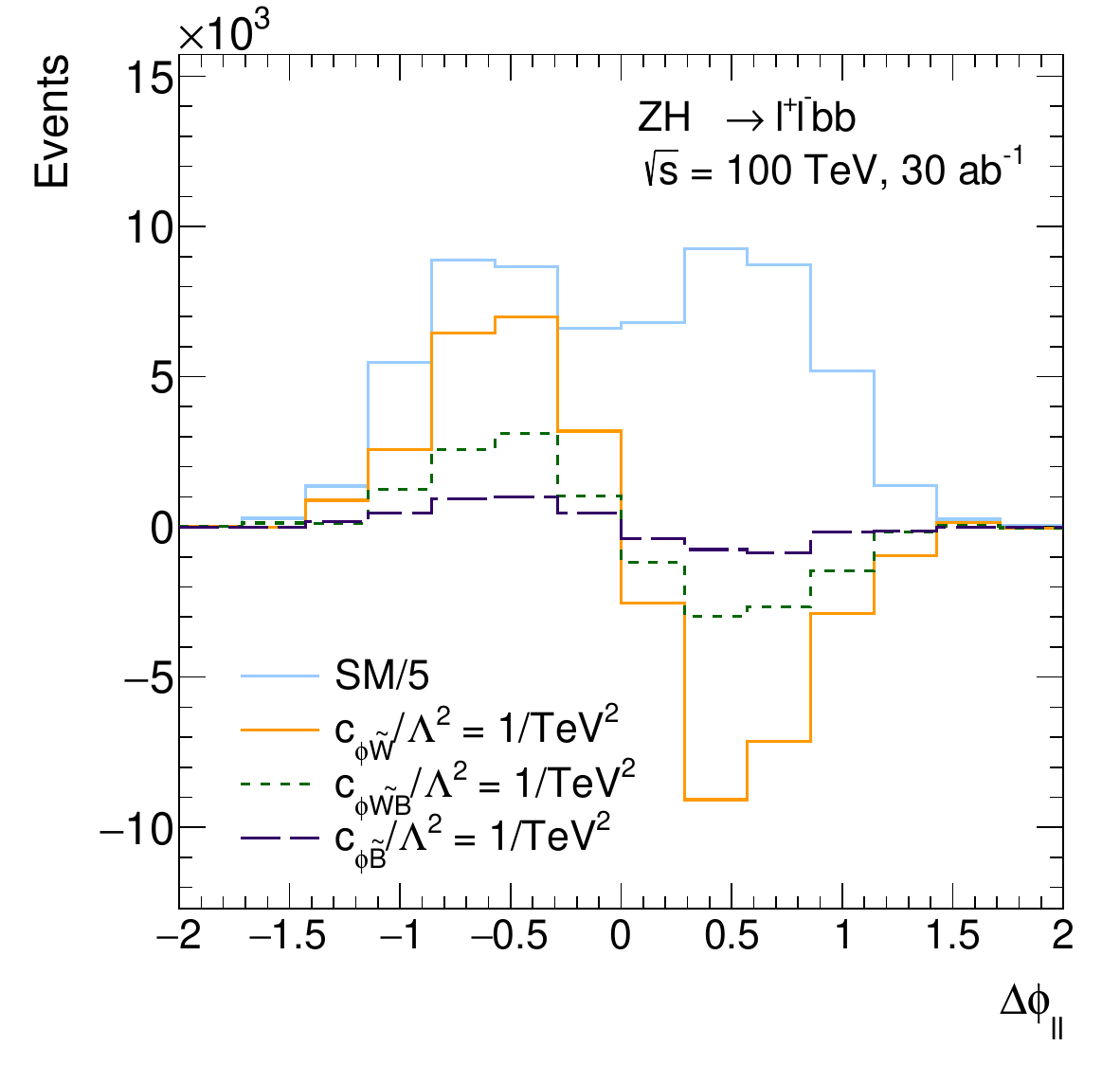}}
  \subfigure[]{\includegraphics[width=0.49\textwidth]{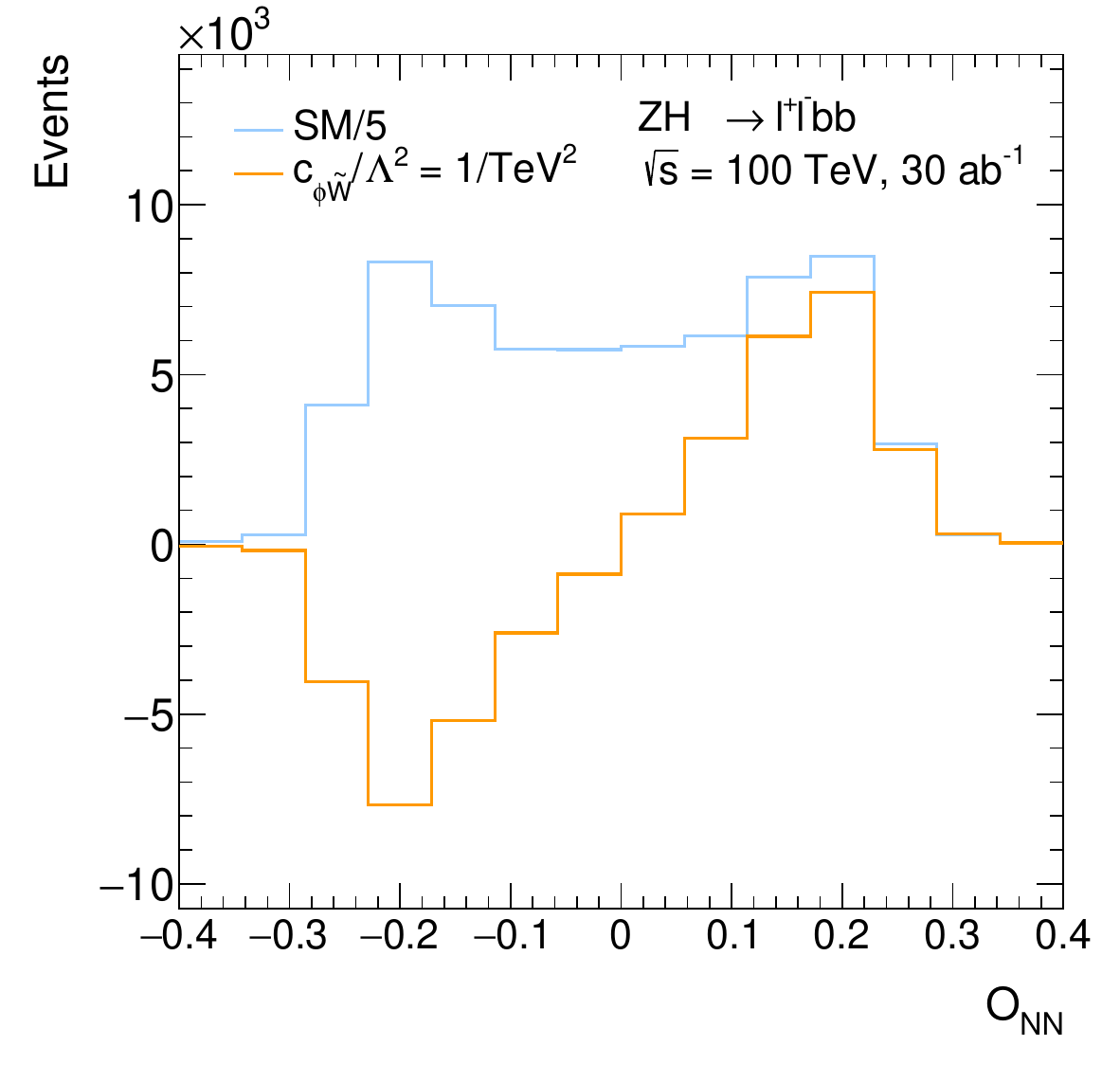}}
  \caption{(a) Expected event yields for $ZH$ production at FCC-hh in the $H\rightarrow b\bar{b}$ decay channel as a function of $\Delta\phi_{\ell\ell}$. (b) Expected event yield as a function of the $O_{\rm NN}^{\rm binary}$ observable using a network trained on the interference contribution induced by the \OHWtil{} operator.
  \label{fig:ppzh_distributions}}
\end{figure}
%%%%%%%%%%%%%%%%%%%%%%%%%%%%%%%%%%%%%%%%%%%%%%%%%%%%%

The sensitivity to CP violation in the $HZZ$ coupling is investigated using the signed azimuthal angle between the two leptons, $\Delta \phi_{\ell\ell}$, as well as observables constructed from ML classifiers. Figure~\ref{fig:ppzh_distributions} shows the expected event yields at FCC-hh as a function of (a) $\Delta\phi_{\ell\ell}$ and (b) the $O_{\rm NN}^{\rm binary}$ observable obtained after training a binary classifier on the interference sample from the \OHWtil{} operator. The use of a binary classifier slightly improves the separation of the SM and interference contributions when compared to $\Delta\phi_{\ell\ell}$ alone. The use of a multiclass network was investigated, but no real improvement in sensitivity was observed due to (i) the lack of statistics in the MC samples and (ii) the fact that the jets merge into a single object at high Higgs boson transverse momentum. Further studies using large-$R$ jets and jet substructure techniques would likely improve sensitivity in this channel further.

The 95\% confidence intervals on the \cHBtil$/\Lambda^2$, \cHWtil$/\Lambda^2$, and \cHWtilB$/\Lambda^2$ Wilson coefficients are presented in Tab.~\ref{tab:ppzh_limits} at LHC, HL-LHC and FCC-hh. The use of a binary classifier improves the constraints by approximately 20-30\% compared to the use of $\Delta\phi_{\ell\ell}$. As expected, FCC-hh performs much better than HL-LHC for measurements in the $ZH$ production channel. Interestingly, for this production and decay channel, the sensitivity to CP-violating $HZZ$ interactions at FCC-hh is comparable to that obtained at FCC-ee for the \OHWtil{} operator. However, FCC-hh is much less sensitive than FCC-ee for the \OHBtil{} and \OHWtilB{} operators.

%\textcolor{red}{REMOVE: Although we do not focus on this process in this work, $W$-associated Higgs production offers another avenue to constrain CP-odd gauge Higgs interactions at $pp$ colliders. In particular, good sensitivity can be obtained to $\OHWtilB{}$ as discussed in~\cite{Bishara:2020vix,Barrue:2023ysk,Rossia:2024rfo}.}

%%%%%%%%%%%%%%%%%%%%%%%%%%%%%%%%%%%%%%%%%%%%%%%%%%%%%
\begin{table}[!t]
\centering
\begin{tabular}{ccccc}
\toprule 
\multirow{2}{*}{Collider} & \multirow{2}{*}{Observable} & \multicolumn{3}{c}{95\% confidence intervals} [TeV$^{-2}$] \\ \cmidrule{3-5}
& & \cHWtil$/\Lambda^2$ & \cHBtil$/\Lambda^2$ & \cHWtilB$/\Lambda^2$  \\
\midrule
\multirow{2}{*}{LHC} & $\Delta\phi_{\ell\ell}$ & [-5.5, 6.0] &  [-50, 50] & [-15, 14] \\
& $O_{\rm NN}^{\rm binary}$ & [-3.8, 3.8] & [-36, 36] & [-9.7, 9.7] \\
\midrule
\multirow{2}{*}{HL-LHC} & $\Delta\phi_{\ell\ell}$ & [-1.2, 1.2]& [-15, 15] & [-4.3, 2.2]\\
& $O_{\rm NN}^{\rm binary}$ & [-0.83, 0.83] & [-7.7, 7.7] & [-2.1, 2.1] \\
\midrule
\multirow{2}{*}{FCC-hh} & $\Delta\phi_{\ell\ell}$ &  [-0.31, 0.31] & [-2.6, 2.6] & [-0.82, 0.82] \\
& $O_{\rm NN}^{\rm binary}$ & [-0.22, 0.22] & [-2.2, 2.2] & [-0.56, 0.56] \\
\bottomrule
\end{tabular}
\caption{95\% confidence intervals on the \cHWtil$/\Lambda^2$, \cHBtil$/\Lambda^2$, and \cHWtilB$/\Lambda^2$ Wilson coefficients obtained from studies of $ZH$ production in the $H\rightarrow b\bar{b}$ decay channel at LHC, HL-LHC and FCC-hh. The CP sensitive observables include the signed azimuthal angle between the two leptons from the $Z$ boson decay, as well as the optimised observables defined using a multiclass NN.}
\label{tab:ppzh_limits}
\end{table}
%%%%%%%%%%%%%%%%%%%%%%%%%%%%%%%%%%%%%%%%%%%%%%%%%%%%%

%%%%%%%%%%%%%%%%%%%%%%%%%%%%%%%%%%%%%%%%%%%%%%%%%%%%%
\section{$H \rightarrow 4\ell$ at $pp$ colliders}
\label{sec:pph4l}
%%%%%%%%%%%%%%%%%%%%%%%%%%%%%%%%%%%%%%%%%%%%%%%%%%%%%
The sensitivity to CP-violating interactions in $H \rightarrow 4\ell$ decays is studied using \mg{} samples of $gg \rightarrow H \rightarrow e^+e^- \mu^+\mu^-$. The cross section of this sample is only valid to leading-order in QCD and is therefore scaled by a factor $\sigma^{\rm N^3LO}_{gg\rightarrow H}/\sigma^{\rm MG}_{gg\rightarrow H}$, where $\sigma^{\rm N^3LO}_{gg\rightarrow H}$ and $\sigma^{\rm MG}_{gg\rightarrow H}$ are the cross sections for gluon fusion production calculated by the LHC Higgs cross section working group \cite{LHCHiggsCrossSectionWorkingGroup:2016ypw} and by \mg{}, respectively. Furthermore, the $H \rightarrow 4\ell$ branching ratio is rescaled to include relevant corrections~\cite{LHCHiggsCrossSectionWorkingGroup:2011wcg}. Concretely, the \mg{} sample cross section is further scaled by a factor $\Gamma^{\rm best}_{H \rightarrow 4\ell}/\Gamma^{\rm MG}_{H \rightarrow 4\ell}$, where $\Gamma^{\rm best}_{H \rightarrow 4\ell}$ and $\Gamma^{\rm MG}_{H \rightarrow 4\ell}$ are the partial widths estimated by the LHC Higgs cross section working group and by \mg{}, respectively. Backgrounds are estimated by producing an inclusive $pp \rightarrow e^+e^- \mu^+\mu^-$ sample, with all SM contributions from $H \rightarrow e^+e^- \mu^+\mu^-$, $Z \rightarrow e^+e^- ,\mu^+\mu^-$ and $ZZ^* \rightarrow e^+e^- \mu^+\mu^-$ included. The background prediction is then estimated per bin of a distribution using this sample, but after subtracting the predicted event yields from the $gg \rightarrow H \rightarrow e^+e^- \mu^+\mu^-$ sample. 

Events are required to pass the selection criteria used by ATLAS for measurements of $H \rightarrow 4\ell$ differential cross sections \cite{ATLAS:2020wny}. Electrons are required to have $p_{\rm T} > 7$~GeV and $|\eta|< 2.47$, whereas muons are required to have $p_{\rm T} > 5$~GeV and $|\eta|< 2.7$. The leptons are then ordered in transverse momentum, and the three leading leptons are required to have $p_{\rm T} > 20,15,10$~GeV, respectively. The leptons are then arranged into same-flavour-opposite-sign (SFOS) pairs, and each pair is required to have $\Delta R > 0.1$. The SFOS pair with invariant mass closest to the $Z$-boson mass is required to have $50<m_{\ell\ell}<106$~GeV. The other SFOS pair is required to have $m_{\rm min} < m_{\ell\ell} < 115$~GeV, where $m_{\rm min}$ is a threshold ranging from 12~GeV to 50~GeV~\cite{ATLAS:2020wny}. Finally, the invariant mass of the four-lepton system is required to be $120<m_{4\ell{}}<130$~GeV.

After applying the selection criteria, 79 SM $H \rightarrow e^+e^- \mu^+\mu^-$ events are predicted for the LHC Run-II dataset. This is similar to the predicted yield obtained in the ATLAS analysis~\cite{ATLAS:2020wny}; the remaining differences can be attributed to the simple detector efficiency and resolution parameterisations used in \delphes{}. The SM Higgs boson event yields predicted at HL-LHC and FCC-hh are 1700 and 364,000, respectively. The background predictions are scaled by a constant normalisation factor in order to reproduce the measured background yields in the ATLAS analysis and account for missing higher-order effects in the QCD calculation. The final predicted background yields are 50, 1070 and 89,000 at LHC Run-II, HL-LHC and FCC-hh, respectively.

%%%%%%%%%%%%%%%%%%%%%%%%%%%%%%%%%%%%%%%%%%%%%%%%%%%%%
\begin{figure}[!t]
  \centering
  \subfigure[]{\includegraphics[width=0.46\textwidth]{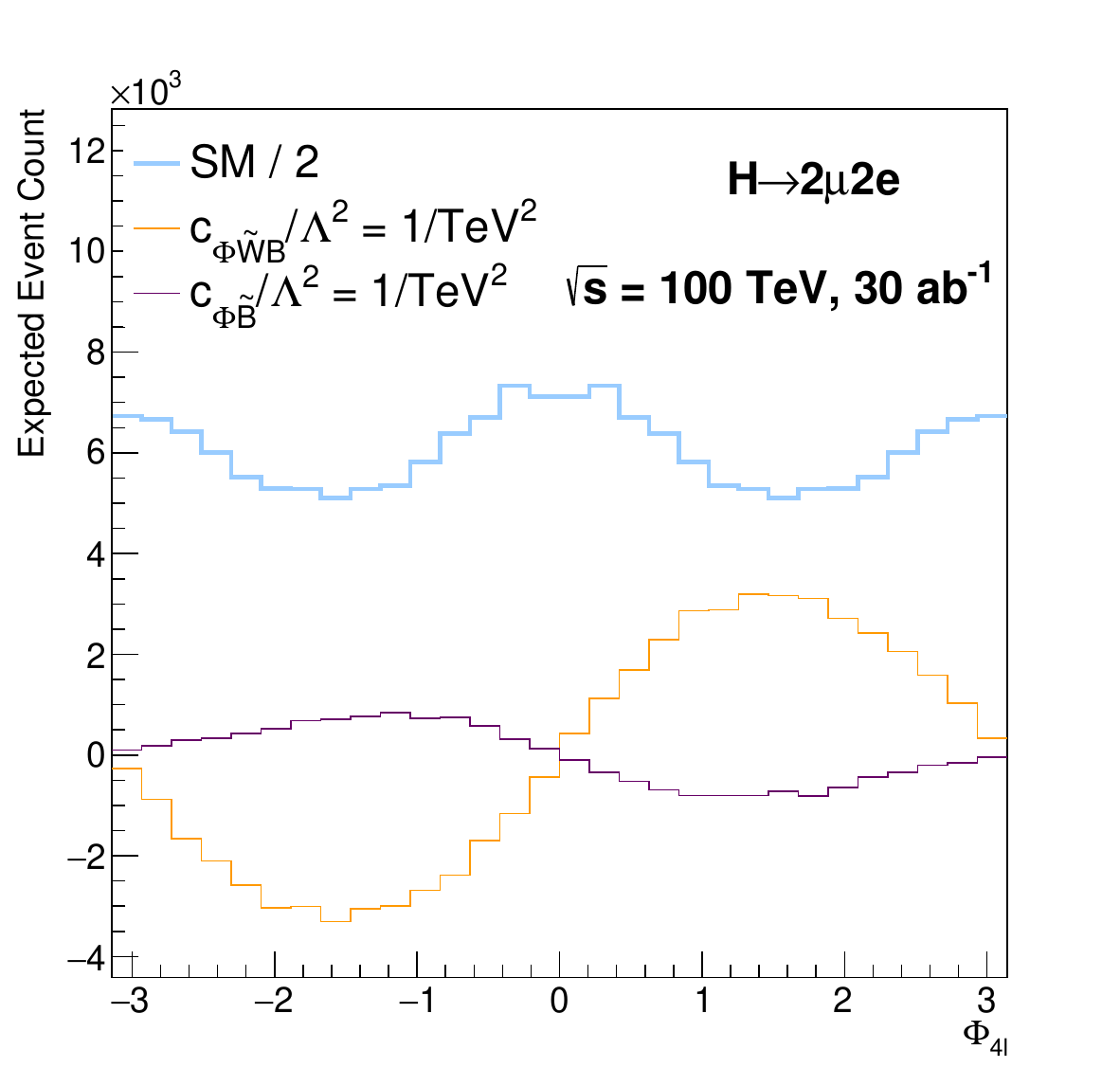}}\quad
  \subfigure[]{\includegraphics[width=0.45\textwidth]{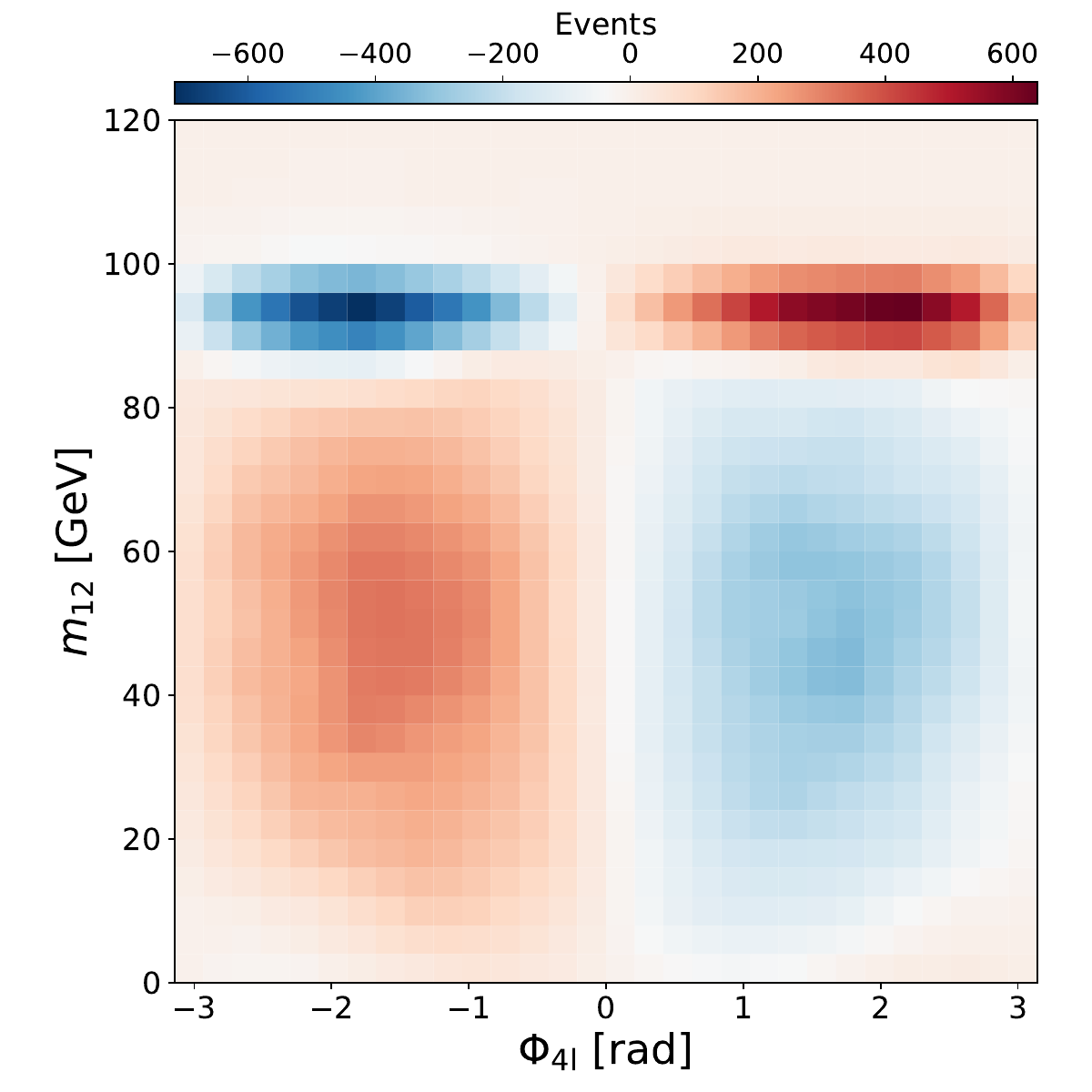}}
  \caption{(a) The expected event yield for $H \rightarrow e^+e^- \mu^+\mu^-$ at FCC-hh as a function of $\Phi_{4\ell}$, for both SM events and the interference contributions predicted for the \OHBtil{} and \OHWtilB{} operators. (b) The expected yield of the interference contribution predicted by the \OHWtilB{} operator as a function of $\Phi_{4\ell}$ and $m_{12}$, where $m_{12}$ is the invariant mass of the SFOS pair that is closest to $m_{Z}$.}
  \label{fig:4l_distributions}
\end{figure}
%%%%%%%%%%%%%%%%%%%%%%%%%%%%%%%%%%%%%%%%%%%%%%%%%%%%%

Figure~\ref{fig:4l_distributions} shows the expected event yield for $H \rightarrow e^+e^- \mu^+\mu^-$ at FCC-hh as a function of (a) $\Phi_{4\ell}$ and (b) as a function of both $\Phi_{4\ell}$ and $m_{12}$, where $m_{12}$ is the invariant mass of the SFOS pair that is closest to the $Z$ pole. The features of the distributions are broadly similar to those observed in previous LHC-based studies~\cite{Bhardwaj:2021ujv}, specifically that the observed asymmetry in $\Phi_{4\ell}$ flips in sign when comparing on-shell and off-shell $Z$ bosons. 

The 95\% confidence intervals on the \cHBtil$/\Lambda^2$ and \cHWtilB$/\Lambda^2$ Wilson coefficients are presented in Tab.~\ref{tab:4l_limits} at LHC Run-II, HL-LHC and FCC-hh. The expected sensitivities at LHC are in good agreement with previous studies~\cite{Bhardwaj:2021ujv}. The much larger event yields at HL-LHC and FCC-hh allow a finer binning of the $\Phi_{4\ell}$ and $m_{12}$ phase space in the likelihood fit, which in turn improves the constraints on the Wilson coefficients by more than naively expected from just cross section and luminosity considerations. The constraints on the \cHBtil$/\Lambda^2$ and \cHWtilB$/\Lambda^2$ Wilson coefficients at FCC-hh are a factor of 20 (150) better than those obtained at HL-LHC (LHC) for this process. When compared to the constraints obtained for $ZH$ production in the $H\rightarrow b\bar{b}$ decay channel at FCC-ee (Sec.~\ref{sec:eezh}), the constraints on the \cHBtil$/\Lambda^2$ and \cHWtilB$/\Lambda^2$ Wilson coefficients are improved by approximately an order of magnitude. 

%%%%%%%%%%%%%%%%%%%%%%%%%%%%%%%%%%%%%%%%%%%%%%%%%%%%%
\begin{table}[!t]
\centering
\begin{tabular}{cccc}
\toprule 
\multirow{2}{*}{Collider} & \multirow{2}{*}{Observable} & \multicolumn{2}{c}{95\% confidence intervals} [TeV$^{-2}$] \\ \cmidrule{3-4}
& & \cHBtil$/\Lambda^2$ & \cHWtilB$/\Lambda^2$ \\
\midrule
\multirow{2}{*}{LHC} & $\Phi_{4\ell}$ & [-1.4, 1.4] & [-5.5, 5.5] \\
& $\Phi_{4\ell}$ vs $m_{12}$ & [-0.83, 0.84] & [-1.7, 1.7] \\
\midrule
\multirow{2}{*}{HL-LHC} & $\Phi_{4\ell}$ & [-0.29, 0.29]& [-1.2, 1.2] \\
& $\Phi_{4\ell}$ vs $m_{12}$ & [-0.14, 0.14] & [-0.27, 0.27] \\
\midrule
\multirow{2}{*}{FCC-hh} & $\Phi_{4\ell}$ & [-0.017, 0.017] & [-0.077, 0.077] \\
& $\Phi_{4\ell}$ vs $m_{12}$ & [-0.007, 0.007] & [-0.015, 0.015] \\
\bottomrule
\end{tabular}
\caption{95\% confidence intervals on the \cHBtil$/\Lambda^2$ and \cHWtilB$/\Lambda^2$ Wilson coefficients obtained using $ H \rightarrow e^+e^-\mu^+\mu^-$ production at the LHC, HL-LHC, and FCC-hh. \cHWtil is predominantly constrained using WBF discussed in Sec.~\ref{sec:vbfh}.}
\label{tab:4l_limits}
\end{table}
%%%%%%%%%%%%%%%%%%%%%%%%%%%%%%%%%%%%%%%%%%%%%%%%%%%%%

%%%%%%%%%%%%%%%%%%%%%%%%%%%%%%%%%%%%%%%%%%%%%%%%%%%%%
\section{$Hjj$ production at $pp$ colliders}
\label{sec:vbfh}
%%%%%%%%%%%%%%%%%%%%%%%%%%%%%%%%%%%%%%%%%%%%%%%%%%%%%
Weak boson fusion production of the Higgs boson has been shown to provide the most stringent constraints on the CP-odd operator \cHWtil~\cite{ATLAS:2023mqy}. As such, the sensitivity to this particular operator at $pp$ colliders was further studied in $Hjj$ production. Decays of the Higgs boson into two tau leptons (\Htautau) are targeted~\cite{Rainwater:1998kj}. A sample of $q\overline{q} \rightarrow \tau^+\tau^- q\overline{q}$ is generated at leading order using \mg{}, for both WBF and $VH$ events. The invariant mass of the final state leptons ($m_{ll}$) of the generated events is required to be within [123, 127]~GeV, in order to ensure that a very large fraction of \Htautau events is produced. The cross section of the generated sample is scaled by a factor of $\sigma^{\rm NNLO}_{\text{WBF}}/\sigma^{\text{MG}}_{\text{WBF}}$, where $\sigma^{\rm NNLO}_{\text{WBF}}$ and $\sigma^{\text{MG}}_{\text{WBF}}$ are the WBF cross section calculated by the LHC Higgs Working group~\cite{LHCHiggsCrossSectionWorkingGroup:2016ypw,Mangano:2017tke} and \mg{}, respectively. An additional $q\overline{q} \rightarrow \tau^+\tau^- q\overline{q}$ sample is generated at leading order with a requirement of 78 $ < m_{ll} < $ 103 GeV, in order to estimate the dominant $Zjj$ background. 

Events are selected following the criteria used by ATLAS for measurements of WBF Higgs boson production in the \Htautau decay channel~\cite{ATLAS:2022yrq}. Three final states are targeted: two different flavour leptonically decaying tau leptons ($\tau_{\text{lep}}\tau_{\text{lep}}$); one hadronically and one leptonically decaying tau lepton ($\tau_{\text{had}}\tau_{\text{lep}}$); and two hadronically decaying tau leptons ($\tau_{\text{had}}\tau_{\text{had}}$). Electrons are required to have $p_{\rm T} > 15$~GeV and $|\eta|< 2.47$, excluding the ATLAS detector transition region of $1.37 < |\eta|< 1.52$. Muons are required to have $p_{\rm T} > 10$~GeV and $|\eta|< 2.5$. $\tau_{\text{had}}$ candidates are required to have $p_{\rm T} > 20$~GeV and $|\eta|< 2.47$, vetoing the transition region. Jets are required to have $p_{\rm T} > 20$~GeV and $|\eta|< 2.5$ and to not be tagged as tau leptons. The impact of ATLAS triggers are emulated by requiring that events have either at least one electron with $p_{\rm T} > 27$~GeV, one muon with $p_{\rm T} > 27.3$~GeV, one electron with $p_{\rm T} > 18$~GeV and one muon with $p_{\rm T} > 14.7$~GeV, or two $\tau_{\text{had}}$ with $p_{\rm T} > 40$~GeV and $p_{\rm T} > 30$~GeV, respectively.

A dedicated event selection is applied depending on the analysis channel considered. In the $\tau_{\text{lep}}\tau_{\text{lep}}$ channel, events are required to have one electron and one muon with an invariant mass of $30 < m_{e\mu} < 100$~GeV. In the $\tau_{\text{lep}}\tau_{\text{had}}$ channel, events are considered if they have one $\tau_{\text{had}}$ and either one electron or one muon. The transverse mass of the lepton and missing transverse momentum system ($m_{\rm T}$) is required to be larger than 70~GeV. Events are selected in the $\tau_{\text{had}}\tau_{\text{had}}$ channel requiring exactly two $\tau_{\text{had}}$ candidates. In all decay channels, the two final state objects are required to have opposite charge and angular requirements are placed on $\Delta R_{\tau\tau}$ and $|\Delta \eta_{\tau\tau}|$ and the $p_{T}$ of the leading jet in the event. The invariant mass of the di-tau system calculated using the collinear approximation is required to satisfy $ m_{\tau\tau}^{\text{coll}} > $ 110 GeV. Furthermore, selection requirements are applied on the missing transverse momentum ($E_{\rm{T}}^{\text{miss}}$) and on the fraction of $p_{\rm T}(\tau)$ carried by the visible decay products. 

%%%%%%%%%%%%%%%%%%%%%%%%%%%%%%%%%%%%%%%%%%%%%%%%%%%%%
\begin{figure}[!t]
  \centering
  \subfigure[]{\includegraphics[width=0.46\textwidth]{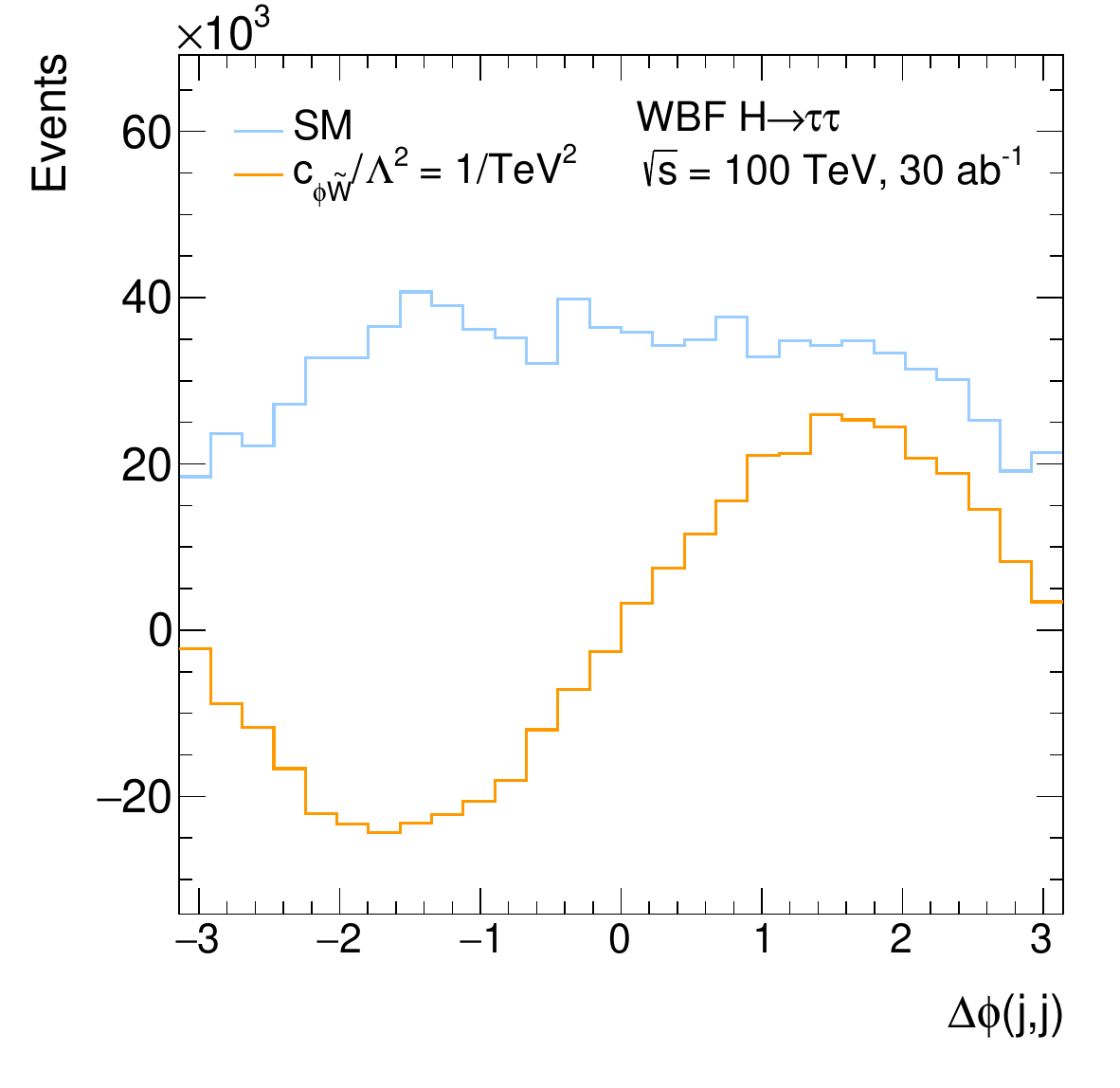}}\quad
  \subfigure[]{\includegraphics[width=0.46\textwidth]{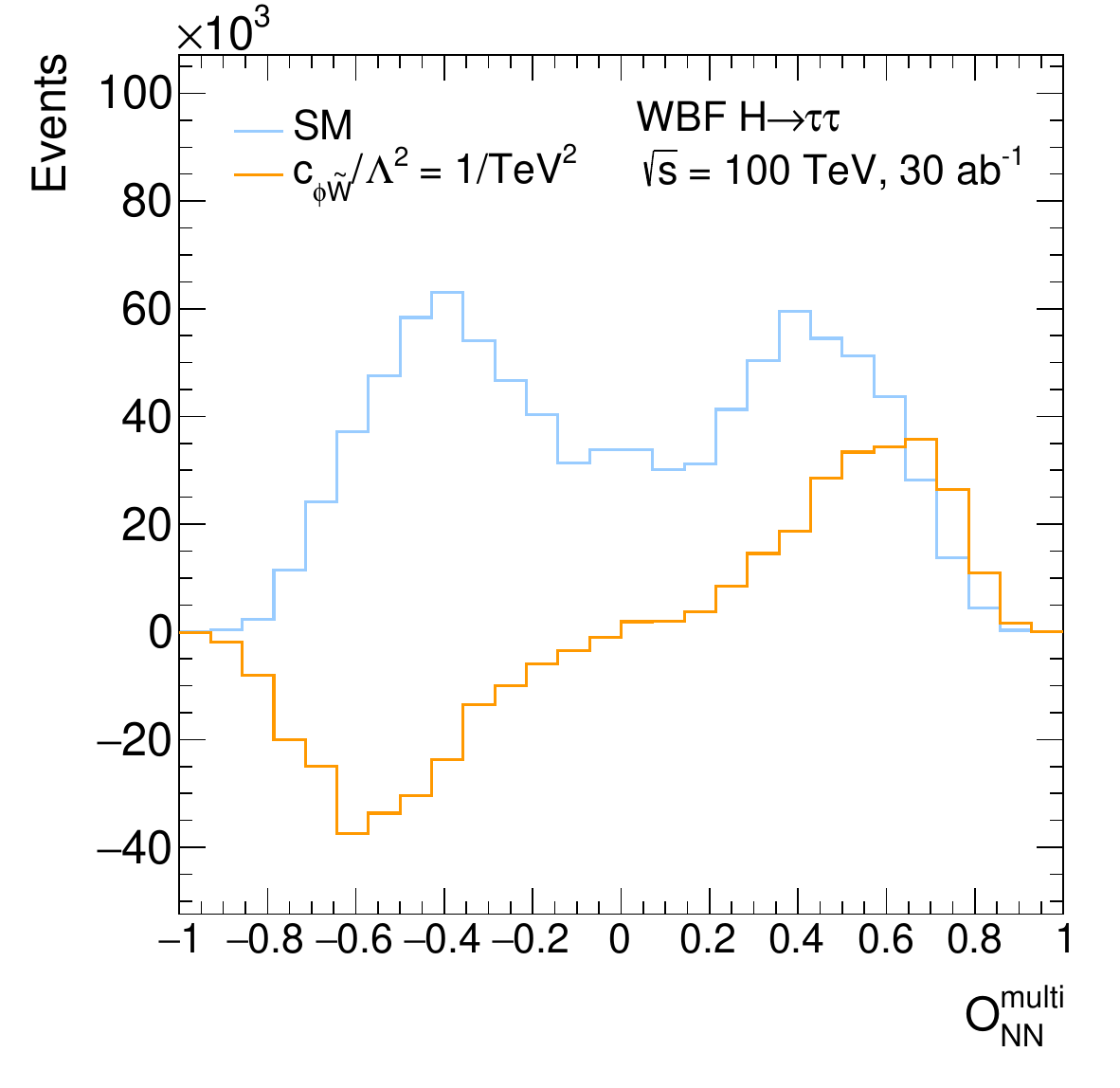}}
  \caption{(a) The expected event yields for $H(\rightarrow\tau\tau)jj$ at FCC-hh as a function of $\Delta\phi_{jj}$, for both SM events and the interference contributions predicted for the \OHWtil{} operator. (b) The same expected yields as a function of the CP-odd observable produced by a multiclass NN.}
  \label{fig:VBF_distributions}
\end{figure}
%%%%%%%%%%%%%%%%%%%%%%%%%%%%%%%%%%%%%%%%%%%%%%%%%%%%%

Events are required to have at least two jets and are vetoed if they have one $b$-tagged jet passing the aforementioned jet object selections. The invariant mass of the two leading jets in the event must be greater than 350~GeV, and the absolute pseudorapidity difference between them must be greater than 3. Furthermore, the two jets must be in opposite hemispheres of the detector, and the tau lepton decay products must be contained in the rapidity interval between those two jets. Finally, the subleading jet must satisfy $p_{\rm T} > 30$~GeV. Following application of the selection requirements described, the SM prediction for the number of $H(\rightarrow\tau\tau)jj$ events produced is 212, 4580 and 862,000 at LHC Run-II, HL-LHC and FCC-hh, respectively. The predicted background yields are 1,680, 36,200 and 16,000,000 at the LHC, HL-LHC and the FCC-hh, respectively.\footnote{The SM yields are validated against the relevant ATLAS publication~\cite{ATLAS:2022yrq}. This analysis uses the same selection criteria used in this study, with the exception of the requirement of $m_{\tau\tau}^{\text{coll}} > $ 110 GeV. Instead, a requirement of  $m_{\tau\tau}^{\text{coll}} > (m_{Z} - 25)$~GeV, where $m_{Z}$ is the mass of the $Z$ boson~\cite{ParticleDataGroup:2024cfk}, is only applied in the lepton-lepton channel. The validation is done against the sum of yields from regions WBF\_0 and WBF\_1 described in Ref.~\cite{ATLAS:2022yrq}, which are defined through the application of a boosted decision tree tagger. The $Hjj$ predictions obtained in this study are very similar to the yields reported by ATLAS. The $Zjj$ prediction is scaled by a factor $N_{\text{total}}/N_{Zjj}$, in which $N_{\text{total}}$ is the total number of background events reported in Ref.~\cite{ATLAS:2022yrq}. A different scaling factor is derived for each final state considered in the analysis. This scaling factor is applied to the $Zjj$ background sample in the studies at all colliders.}

Figure~\ref{fig:VBF_distributions} shows the expected event yields for $H(\rightarrow\tau\tau)jj$ at FCC-hh as a function of (a) $\Delta\phi_{jj}$ and (b) as a function of the CP-odd observable produced by a multiclass NN. The NN was trained on jet kinematics using both the interference event sample and the SM $Hjj$ event sample. It was verified that the inclusion of kinematic information from the di-tau system provides no significant improvement to the NN performance. As expected, for both CP-odd observables, the interference contribution is asymmetric around $\Delta\phi_{jj}=0$ and integrates to 0, whereas the pure-SM distribution is symmetric. 

\begin{table}[!t]
\centering
\begin{tabular}{ccc}
\toprule
Collider & Observable & $\cHWtil$/$\Lambda^2$[TeV$^{-2}$] \\
\midrule
\multirow{2}{*}{LHC} & $\Delta\phi_{jj}$ & [-0.89, 0.89]  \\
& $O_{\rm NN}^{\text{multi}}$ & [-0.74, 0.73]  \\
\midrule
\multirow{2}{*}{HL-LHC} & $\Delta\phi_{jj}$& [-0.19, 0.19] \\
& $O_{\rm NN}^{\text{multi}}$ & [-0.16, 0.16] \\
\midrule
\multirow{2}{*}{FCC-hh} & $\Delta\phi_{jj}$ & [-0.016, 0.016]  \\
& $O_{\rm NN}^{\text{multi}}$ & [-0.007, 0.007]  \\
\bottomrule
\end{tabular}
\caption{95\% confidence intervals on the \cHWtil$/\Lambda^2$ Wilson coefficient, obtained from analysis of WBF Higgs production in the $H\rightarrow \tau^+ \tau^-$ decay channel at the LHC, HL-LHC and FCC-hh.}
\label{tab:VBF_limits}
\end{table}

WBF is predominantly sensitive to \OHWtil~in contrast to $H\to ZZ$ discussed in the previous section, and we focus on this interaction in the following. The 95\% confidence intervals on the $\cHWtil/\Lambda^2$ Wilson coefficient for the different $pp$ colliders in study are presented in Tab.~\ref{tab:VBF_limits}, using either $\Delta\phi_{jj}$ or $O_{NN}$ for the statistical analysis. The expected sensitivities at the LHC are similar to recent constraints by ATLAS on $\cHWtil/\Lambda^2$, set using the same signal process~\cite{ATLAS:2024wfv}. They are also in good agreement with previous studies, which investigated the use of the same NN-based observables~\cite{Bhardwaj:2021ujv}. The expected sensitivity to the \OHWtil{} operator at FCC-hh is improved by a factor of 20 (100) compared to HL-LHC (LHC). Furthermore, this channel provides the best overall sensitivity to the \OHWtil{} operator, when compared to the constraints obtained for other production and decay channels considered at FCC-hh (Secs.~\ref{sec:ppzh}~and~\ref{sec:pph4l}).
When compared to the constraints obtained for $ZH$ production in the $H\rightarrow b\bar{b}$ decay channel at FCC-ee (Sec.~\ref{sec:eezh}), the constraints on $\cHWtil/\Lambda^2$ are improved by an order of magnitude.

%%%%%%%%%%%%%%%%%%%%%%%%%%%%%%%%%%%%%%%%%%%%%%%%%%%%%
\section{Discussion}
\label{sec:disc}
%%%%%%%%%%%%%%%%%%%%%%%%%%%%%%%%%%%%%%%%%%%%%%%%%%%%%
The results presented in Sections~\ref{sec:eezh}--\ref{sec:vbfh} demonstrate two clear features. First, the constraints on CP-violating $HZZ$ interactions will be improved at all future colliders with respect to the HL-LHC alone. Second, the constraints obtained at a future hadron collider operating at $\sqrt{s}=100$~TeV will be much tighter than those obtained at a future electron-positron collider operating at $\sqrt{s}=240$~GeV; this implies that the FCC-hh is better suited to studying CP-violation in the Higgs sector. In this section, we consider various limitations that are present in our studies to demonstrate that the main conclusions of the paper are robust. 

In our analysis, constraints are placed on Wilson coefficients assuming that a single operator describes any new physics effects that are present in the data. A more realistic and model-independent assumption would be to allow all three operators to be simultaneously constrained in the fit. The impact of this approach would be to weaken the constraints and induce blind directions in the parameter space, due to the cancellation of asymmetries induced by different operators. These blind directions can, however, be limited by performing a global fit across all available CP-sensitive measurements~\cite{ATLAS:2026urg}, because each measurement has different sensitivity to each operator. In a global fit approach, all new measurements from a future collider would be combined with all previous measurements at HL-LHC. 

The operators that induce CP-asymmetries in $HZZ$ interactions also affect other final states, and measurements of these final states would provide additional sensitivity in a global fit. Measurements of $WH$ production~\cite{Bishara:2020vix,Barrue:2023ysk,Rossia:2024rfo}, inclusive $W\gamma$ production~\cite{ATLAS:2026xuq} and VBF $Z$ production~\cite{ATLAS:2020nzk} at hadron colliders would add substantial extra sensitivity. This also extends to dimension-six operators not considered in this work, e.g.~\cite{DasBakshi:2020ejz,Biekotter:2021int}. Likewise, measurements of $H\ell^+\ell^-$ production at additional centre-of-mass energies at a lepton collider could help untangle the contributions from $ZH$ production and VBF Higgs production. Furthermore, it is important to recognise that the operators that induce CP-asymmetries in $HZZ$ interactions also affect non-collider experiments, and CP-violation has been formidably constrained by the measurement of anomalous electric dipole moments (EDMs). The EDM constraints, however, are also susceptible to blind directions in the parameter space and are particularly affected when CP-violation across the three fermion families is considered~\cite{Brod:2022bww}. Global fits to collider and EDM measurements will mitigate this issue and additional flavour-specific insight can be gained from collider-based analyses of top quark interactions~\cite{Englert:2019xhk,Bhardwaj:2023ufl}.

At hadron colliders, the validity of the EFT expansion is a valid concern because the large centre-of-mass energy provides access to momentum scales at (or above) the new physics scale~$\Lambda$. Such considerations have limited impact on the results presented in this paper. First, the $H\rightarrow4\ell$ measurements are performed at the Higgs pole and therefore largely unaffected. Second, for $ZH$ production, questions over EFT validity would only arise if the $Z$ and $H$ have high momentum. However, this coincides with the `boosted' limit in which the Higgs boson decay products are merged into a single large-radius jet; our results are obtained for resolved jets at lower transverse momentum. Finally, for VBF Higgs production, we assessed the EFT validity directly by rederiving constraints on Wilson coefficients after removing EFT contributions that are above a given momentum scale. The constraints on the Wilson coefficients were unchanged if the jet momenta were restricted to $p_{\rm T}<500$~GeV and weakened only by a factor of 1.4 if the dijet invariant mass was restricted to $m_{jj}<2$~TeV. 

Finally, our analysis is limited to tree-level sensitivity explorations. It is, however, worthwhile mentioning that the expected sensitivity at $e^+ e^-$ colliders will enable a detailed exploration of radiative imprints of CP violation. For instance, interference with the imaginary part of the one-loop amplitude with the CP-violating operators will then manifest itself in asymmetries of CP-even observables, e.g., in the $Z$ boson's transverse momentum~\cite{Asteriadis:2024xuk,Asteriadis:2024xts}.\footnote{We thank Sally Dawson and Pier Paolo Giardino for bringing this to our attention.} Formally, these effects are suppressed by a loop order compared to the tree-level analysis presented in this work. The high precision that can be obtained, particularly at an FCC-ee, however, opens up this direction as a complementary way for studying CP violation, which requires careful disentangling from other competing effects. In the SM, the only source of CP-violation is the phase of the CKM matrix, which has a small relevance compared to residual theoretical uncertainties (in practical high-energy calculations at proton machines, the quark mixing is assumed to be absent to a very good approximation~\cite{Denner:1991kt}). Higgs production $e^+e^-\to ZH$ process in the SM does not have a dependence on the CKM matrix at NLO, as only leptonic charged currents are probed in this process. At hadron colliders, the dominant uncertainty arises from QCD corrections. As QCD is CP-conserving, the differential asymmetries considered here are robust beyond rate changes. The QCD corrections to WBF are known to be small, e.g.~\cite{Figy:2003nv}. Regions of electroweak Sudakov enhancement~\cite{Ciccolini:2007ec,Ciccolini:2007jr,Figy:2010ct} do not drive the sensitivity (see above); we therefore expect our results to be robust after rescaling to total cross sections.

%%%%%%%%%%%%%%%%%%%%%%%%%%%%%%%%%%%%%%%%%%%%%%%%%%%%%
\section{Summary and Conclusions}
\label{sec:conc}
%%%%%%%%%%%%%%%%%%%%%%%%%%%%%%%%%%%%%%%%%%%%%%%%%%%%%
The absence of concrete evidence for new physics beyond the Standard Model at the energy frontier explored by the Large Hadron Collider contrasts sharply with experimentally established phenomena that the SM fails to explain. This apparent conundrum underscores the need for precision studies beyond the HL-LHC era, especially if no new physics signals emerge there. One such area is the search for additional CP violation, which, through Sakharov's criteria, can be linked to the observed baryon asymmetry of the universe as part of electroweak baryogenesis. While the HL-LHC will significantly enhance sensitivity to CP-violating effects, future experimental programmes at electron-positron and proton-proton colliders have the potential to set even more ambitious benchmarks for these searches.

In this work, we present a comprehensive comparison of sensitivity to new sources of weak gauge-Higgs CP violation. Leveraging well-motivated, experimentally accessible processes and specific CP-sensitive observables, we provide a detailed outlook on the future gauge-Higgs CP profile, highlighting the capabilities of both electron-positron colliders and proton-proton colliders. As a common theme, we find improvements in sensitivity to CP-violating Higgs boson interactions at all future colliders, when compared to the HL-LHC expectations. The electron-positron colliders (FCC-ee and LCF) are a priori capable of providing a detailed picture of the electroweak Higgs boson interactions in a clean environment. However, the study of asymmetries in appropriately constructed CP-sensitive observables will reduce the impact of experimental and theoretical systematic uncertainties. This allows FCC-hh to fully utilise the abundance of final states produced in $pp$ collisions, the larger event yields, and the additional kinematic information, which collectively compensate for the busier background environment. Overall, this suggests the FCC-hh is more suited to analyse the Higgs boson CP properties in great detail. 

%%%%%%%%%%%%%%%%%%%%%%%%%%%%%%%%%%%%%%%%%%%%%%%%%%%%%
\subsection*{Acknowledgments}
%%%%%%%%%%%%%%%%%%%%%%%%%%%%%%%%%%%%%%%%%%%%%%%%%%%%%
J.B. is funded by the Science Technology and Facilities Council (STFC) under grant ST/Y509814/1.
A.J.C. is supported by the Leverhulme Trust under grant RPG-2020-004. 
C.E.'s work is supported by the Institute for Particle Physics Phenomenology Associateship Scheme. 
S.F. is supported by the STFC under the grant ST/W000482/1 and the European Research Council (ERC) under the European Union’s Horizon 2020 research and innovation programme grant agreement 772357-Open3Gen.
J.N is funded by the Royal Society under grant URF/R/241013.
L.S.P.T is funded by the DESY laboratory.
A.P. is supported by the STFC under grant ST/W000601/1 and by the Leverhulme Trust under grant RPG-2020-004.
A.R. is supported by the STFC under grant ST/W000520/1.
J.C.S. is supported by the STFC under grant ST/X005984/1.
S.L.W. is supported by the STFC under grant ST/W00044X/1.   
Y.Z. is supported by the Chinese Scholarship Council under grant 202308090795.

%%%%%%%%%%%%%%%%%%%%%%%%%%%%%%%%%%%%%%%%%%%%%%%%%%%%%
%  bibliography
%%%%%%%%%%%%%%%%%%%%%%%%%%%%%%%%%%%%%%%%%%%%%%%%%%%%% 
%\bibliographystyle{JHEP}
\bibliography{paper.bbl}
%%%%%%%%%%%%%%%%%%%%%%%%%%%%%%%%%%%%%%%%%%%%%%%%%%%%%   
\end{document}